\documentclass[letterpaper,twocolumn,10pt]{article}

\usepackage{usenix}
\usepackage{pdflscape}
\usepackage{graphicx}
\usepackage{pifont} 
\usepackage{rotating}
\usepackage{multirow}
\usepackage{makecell}
\usepackage{comment}
\usepackage{changepage}
\usepackage{xspace}
\usepackage{tabto}
\usepackage{acronym}
\usepackage{tabularx}
\usepackage{paralist}
\usepackage{listings}
\usepackage{xcolor}
\usepackage{colortbl}
\usepackage{xurl}
\usepackage{enumitem}
\usepackage{caption}
\usepackage{amsfonts}
\usepackage{booktabs}
\usepackage{hyperref}
\usepackage{pifont}
\usepackage{tcolorbox}
\usepackage{enumitem}
\usepackage{tikz}
\usepackage{bm}
\usepackage{amsmath}
\usepackage{array} 

\newtcolorbox{fancyquote}[1][]{colback=gray!10, colframe=black!50, sharp corners, fonttitle=\bfseries, boxrule=0.6pt, coltitle=black, title=}

\newcommand{\ie}{i.\@\,e.\@\xspace}
\newcommand{\eg}{e.\@\,g.\@\xspace}

\def\checkonly{\tikz\draw[scale=0.4,fill=black] (0,.35) -- (.25,0) -- (1,.7) -- (.25,.15) -- cycle;}

\begin{document}

\title{Security Analysis of 5G NR Device-to-Device Sidelink Communications}

\author{
{\rm Evangelos Bitsikas}\\
bitsikas.e@northeastern.edu\\
Northeastern University
\and
{\rm Aanjhan Ranganathan}\\
aanjhan@northeastern.edu\\
Northeastern University
}

\maketitle

\begin{abstract}
5G NR sidelink communication enables new possibilities for direct device-to-device interactions, supporting applications from vehicle-to-everything (V2X) systems to public safety, industrial automation, and drone networks. However, these advancements come with significant security challenges due to the decentralized trust model and increased reliance on User Equipment (UE) for critical functions like synchronization, resource allocation, and authorization. This paper presents the first comprehensive security analysis of NR V2X sidelink. We identify vulnerabilities across critical procedures and demonstrate plausible attack, including attacks that manipulate data integrity feedback and block resources, ultimately undermining the reliability and privacy of sidelink communications. Our analysis reveals that NR operational modes are vulnerable, with the ones relying on autonomous resource management (without network supervision) particularly exposed. To address these issues, we propose mitigation strategies to enhance the security of 5G sidelink communications. This work establishes a foundation for future efforts to strengthen 5G device-to-device sidelink communications, ensuring its safe deployment in critical applications.
\end{abstract}

\section{Introduction}

5G NR sidelink communication~\cite{bell-labs, 3gpp.23.287, 3gpp.37.985, 3gpp.33.536, 3gpp.24.587, weerackody23:whoneeds} is a promising technology that enables direct, low-latency device-to-device communication without routing traffic through the core network or a base station. Initially developed to support automotive vehicle-to-everything (V2X) systems, 5G NR sidelink is now proving valuable in a diverse range of applications~\cite{qualcomm2}, including public safety, industrial automation, mission-critical services, proximity-based services, and IoT networks. For example, in V2X networks~\cite{qualcomm3}, sidelink enables vehicles to exchange real-time information, which is crucial for collision avoidance and traffic management. Similarly, in mission-critical applications like drone communication and public safety networks, sidelink promises a resilient communication channel that remains operational even when network infrastructure is compromised. Industry manufacturers~\cite{qualcomm, qualcomm2, bell-labs} have highlighted that the importance of sidelink extends beyond these specific scenarios, as it also enhances network coverage and capacity by offloading traffic from the core network and enabling proximity-based services. In situations where network infrastructure is sparse or compromised, such as in rural areas or disaster zones, sidelink can provide essential communication capabilities.

Despite its numerous advantages, sidelink communication introduces new security concerns that differ from those in traditional cellular networks. Unlike traditional cellular networks where trust is centralized in the network infrastructure, sidelink relies on user-based entities that can be exploited or may include attackers who are valid network subscribers. In sidelink, User Equipment (UE) devices assume greater responsibilities, such as establishing direct connections, extending network coverage, and managing resources -- tasks that \textit{implicitly place increased trust on them}. Many attacks~\cite{Rupprecht18:sok, eleftherakis24:sok} in the cellular ecosystem focus on exploiting the UE, as it is often the most targeted entity due to its accessibility and potential vulnerabilities. With this shift from centralized to distributed trust model in 5G NR sidelink technology, traditional security paradigms no longer suffice. The potential for vulnerabilities and its impact expands, particularly in applications where public safety is on the line. Imagine a scenario in which an attacker disrupts communications between autonomous vehicles, blocking or spoofing messages that are critical for collision avoidance. In such a context, the consequences of a security breach could escalate from network disruption to real-world damage and even loss of life. Without robust security mechanisms, these vulnerabilities could be exploited, causing dangerous communication lapses in contexts where reliability is non-negotiable such as in connected and autonomous transportation systems.

This paper provides a comprehensive analysis of the critical protocol mechanisms and security procedures in 5G V2X communications, systematically identifying unique security challenges by scrutinizing the latest 3GPP specifications (Releases 17-18). Our motivation stems from the fact that the security of 5G sidelink is severely underexplored, especially in terms of low-level physical and MAC-layer vulnerabilities, synchronization, resource allocation and PC5 protection. Most existing studies focus on broad NR V2X risk assessments, often addressing general vehicular communication threats rather than the unique security challenges introduced by cellular-specific mechanisms. It should be acknowledged, however, that the challenge of investigating sidelink is further compounded by the lack of accessible and reliable experimental setups and implementations, reinforcing our motivation for a thorough analysis to advance this area of research. To the best of our knowledge, this is the first technical study to provide a comprehensive, specification-driven security analysis of the 5G V2X sidelink internals. As commercial NR V2X sidelink implementations are expected in the near future, this work will be essential for anticipating and addressing vulnerabilities before widespread real-world deployments, and becomes a stepping stone for future security analyses and testing. 

Specifically, we make the following contributions:
\begin{compactenum}
    \item We provide a comprehensive analysis of the 3GPP specifications and offer a detailed overview of crucial physical-layer and security procedures within 5G NR V2X communications.
    \item Through a rigorous evaluation of the security aspects of sidelink, we identify vulnerabilities that pose significant risks to the integrity, confidentiality, and availability of network communications. These vulnerabilities span critical areas such as synchronization, authorization, broadcast transmission, feedback mechanisms, resource allocation, direct communication messages (PC5), and security parameterization.
    \item Based on these identified vulnerabilities, we design several attacks that exploit flaws in all critical procedures of sidelink. One example is the HARQ (Hybrid Automatic Repeat reQuest) feedback spoofing -- a technique whereby an attacker can inject false feedback to manipulate retransmission behavior and degrade network performance. Together, these attacks demonstrate how malicious actors could disrupt critical UE-to-UE communications in vehicular networks and drone-based systems. 
    \item Finally, we propose a comprehensive set of countermeasures and mitigation strategies, encompassing technical measures, protocol enhancements, and best practices for secure implementation and deployment. We also discuss topics related to false base stations, GNSS attacks, and insider threat that can significantly impact the sidelink network.
\end{compactenum}

As part of our \textbf{ethical and responsible disclosure}, we have completed the GSMA vulnerability disclosure process. GSMA has verified all the findings and assigned the public identifier~\href{https://www.gsma.com/solutions-and-impact/technologies/security/gsma-mobile-security-research-acknowledgements/}{CVD-2024-0098} (TBA).

%a public CVD-XXXX-XXXX~\footnote{anonymized for the review} identification.

\section{Background and Threat Model} \label{sec:prelims}

\subsection{5G NR Sidelink Architecture and Operational Modes}

\begin{figure}[!t]
     \centering
     \includegraphics[width=\columnwidth]{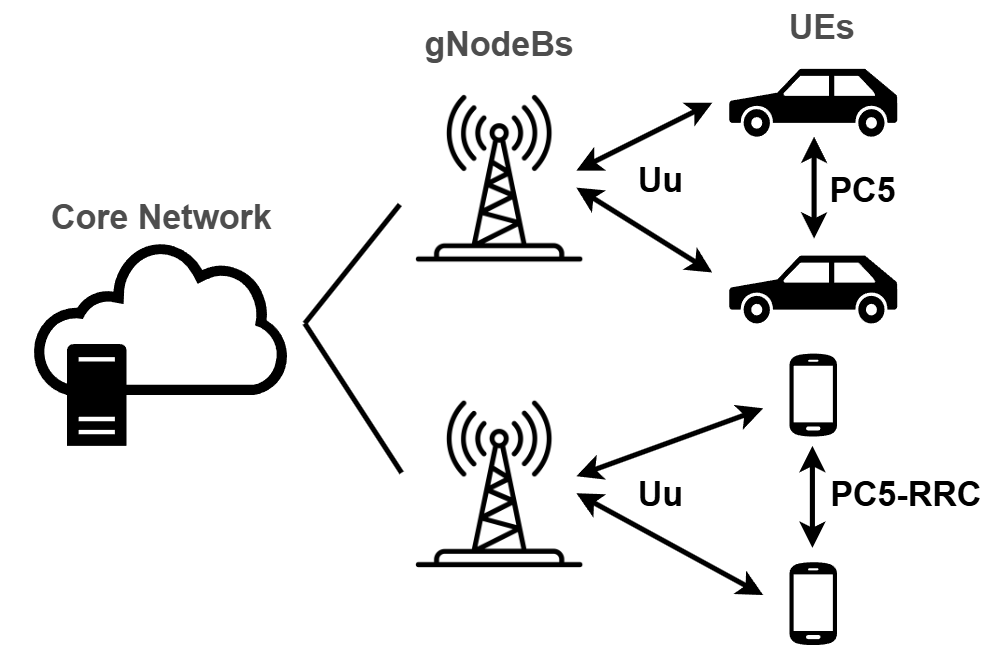}
     \caption{In the 5G architecture, the sidelink components include the UEs communicating over the PC5 or PC5-RRC interface, while potentially having a Uu connection with the network, if available and permitted.}
     \label{fig:architecture}
\end{figure}

In the 5G V2X architecture~\cite{3gpp.23.287, 3gpp.38.331}, the network components and protocols have been adapted to support direct device-to-device (D2D) communications. According to~\cite{3gpp.38.331}, the general D2D communication is simply denoted as NR sidelink, while the vehicle-based (more specialized) is referred to as NR V2X sidelink, which the scenario we adopt for this work. Figure~\ref{fig:architecture} shows the ecosystem of the general and vehicle-based 5G sidelink. 

The core network contains entities like, the \textit{Access and Mobility Management Function (AMF)} which manages the UE registration, access control, mobility, and critical security functions. The \textit{Next-Generation Node B (gNodeB)} connect UEs to the core, and manage radio resource allocation, and synchronize communications. Each \textit{User Equipment (UE)} participates in D2D communication using the PC5 interface and can be a Synchronization Reference (SyncRef). Section~\ref{sec:protocol} provides an overview on the protocol stack used in this architecture.

Based on the specifications~\cite{3gpp.37.985, 3gpp.38.300}, 5G NR sidelink operates in two primary modes: 

\emph{Mode 1: Network-Scheduled.} The gNodeB centrally controls and schedules radio resources for sidelink communication. A UE requests sidelink resources from the network, and the gNodeB allocates specific time-frequency resources based on its scheduling policies and currently available resources. This mode is ideal when UEs are within network coverage and require reliable communication with strict Quality of Service. 

\emph{Mode 2: UE-Autonomous.} UEs manage their own radio resources for sidelink communication without relying on the gNodeB for scheduling decisions. UEs independently choose resources from a predefined sidelink resource pool, either pre-configured at the Mobile Equipment (ME) and/or the Universal Integrated Circuit Card (UICC), containing available time-frequency slots for sidelink transmissions. UEs perform a sensing procedure to identify unoccupied resources and selects its transmission resources to minimize collisions with other UEs. This mode is useful when UEs are outside network coverage, such as in rural or when connectivity is limited.

\subsection{Physical-Layer Channels and Key Mechanisms}

The 5G NR sidelink physical layer relies on specific channels~\cite{3gpp.38.211, 3gpp.38.212, 3gpp.38.213, 3gpp.38.214, 3gpp.37.985, 3gpp.38.133, 3gpp.38.331} that implement crucial mechanisms for synchronization, resource allocation, and the HARQ process. Table~\ref{tab:phy} gives an overview of the physical channels. These radio resource management and control mechanisms are essential to ensure reliable, low-latency for direct device interactions.

\textbf{Synchronization.} Synchronization is critical for coordinating transmissions between UEs, especially in Mode 2 (UE-Autonomous). This function is handled by the Physical Sidelink Broadcast Channel (PSBCH), which broadcasts the Sidelink Synchronization Signal (S-SSB), containing the Primary and Secondary Sidelink Synchronization Signals (S-PSS and S-SSS) along with the Sidelink Synchronization Signal Identifier (SLSS ID). This identifier helps UEs in out-of-coverage scenarios synchronize with a nearby SyncRef UE, ensuring coherent timing and frequency alignment without relying on a base station or GNSS.

\textbf{Resource Allocation.} Resource allocation determines how UEs access the radio spectrum over the sidelink. In Mode 1 is a network-scheduled scheme, where the gNodeB centrally allocates time-frequency resources. In Mode 2, UEs independently select resources from a predefined sidelink resource pool. Resource allocation control can be transmitted via the Physical Sidelink Control Channel (PSCCH), while UEs in Mode 2 also use sensing algorithms to identify unoccupied resources, thereby minimizing interference and optimizing resource use in high-density environments.

\textbf{Integrity and Reliability.} The HARQ (Hybrid Automatic Repeat reQuest) process is crucial for ensuring data integrity and reliability in 5G NR sidelink communication. The Physical Sidelink Shared Channel (PSSCH) carries user data and incorporates HARQ feedback to enable error correction. When errors are detected, the Physical Sidelink Feedback Channel (PSFCH) transmits HARQ acknowledgments (ACKs) and negative acknowledgments (NACKs), allowing UEs to request retransmissions, which is essential for maintaining ultra-reliable low-latency communication (URLLC), especially in applications requiring high levels of data accuracy and reliability.

\begin{table}[t!]
\centering
\caption{5G NR Sidelink Physical Channels}
\begin{tabular}{|l|m{5.5cm}|}
\hline
\textbf{Channel} & \textbf{Purpose} \\
\hline
PSCCH & Manage control info, like scheduling and resource allocation. \\
\hline
PSSCH & Transmit user-plane data and support HARQ for error correction. \\
\hline
PSBCH & Broadcast synchronization info for timing and frequency alignment. \\
\hline
PSFCH & Provide HARQ feedback (ACKs/NACKs) for reliable communication. \\
\hline
\end{tabular} \label{tab:phy}
\end{table}

\subsection{Sidelink Protocol Stack} \label{sec:protocol}

\begin{figure}[!t]
     \centering
     \includegraphics[width=\columnwidth]{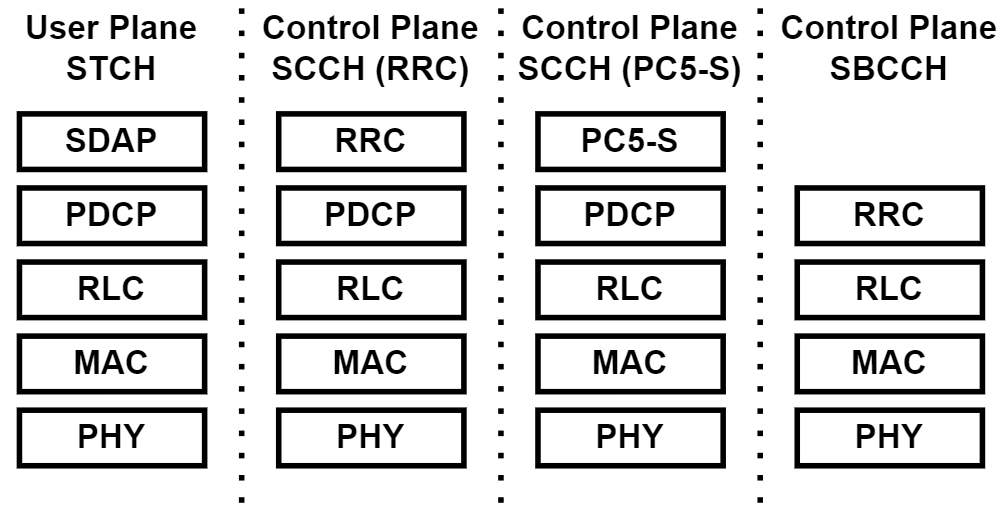}
     \caption{Stack protocols in the PC5 interface.}
     \label{fig:protocols}
\end{figure}

Figure~\ref{fig:protocols} shows the stack protocols in user and control planes for the logical channels in the Proximity Communication 5 (PC5) interface, based on~\cite{3gpp.37.985}. PC5 is the direct communication interface used between two User UEs without the need for network infrastructure.

The protocol stack for the user-plane Sidelink Traffic Channel (STCH) on the PC5 interface includes the Service Data Adaptation Protocol (SDAP), Packet Data Convergence Protocol (PDCP), Radio Link Control (RLC), Medium Access Control (MAC), and the Physical layer. These layers are responsible for handling data transmission between UEs over the direct communication link. For the control plane on the PC5 interface, the AS protocol stack used for Signaling Control Channel (SCCH) related to Radio Resource Control (RRC) signaling consists of the RRC layer, PDCP, RLC, MAC sublayers, and the Physical layer. This stack manages the control messaging required for establishing, maintaining, and releasing connections over the PC5 interface.

Alternatively, the control plane protocol stack for the PC5-S interface, which facilitates control communications, is positioned above the PDCP, RLC, and MAC sublayers, with the Physical layer at the end. These layers ensure reliable control messaging and coordination between devices over the PC5 interface. Finally, the AS protocol stack for the System Broadcast Control Channel (SBCCH) on the PC5 interface comprises the RRC, RLC, and MAC sublayers, along with the Physical layer. This stack is used for broadcasting system information to devices within the communication range.

\subsection{Threat Model}

To assess the security of 5G NR sidelink, we consider an adversary capable of wirelessly intercepting, modifying, and forwarding messages between benign participants (UEs and network entities) over the public channels \textit{of NR V2X environment specifically}. The adversary can deploy malicious UEs and can be a network subscriber in Mode 1 with a valid USIM/eSIM (to collect network parameters and configurations) within the target network to disrupt sidelink communication. The adversary may impersonate a legitimate UE acting as a SyncRef to mislead other UEs, potentially causing disruptions in synchronization, resource allocation, and HARQ processes, which are critical to reliable sidelink operation. However, the adversary does not have physical access to tamper with the SIM card, UE hardware, base station, or core network components and obtain sensitive information, such as cryptographic session keys. In our work, we also consider side-channel attacks, signal jamming attacks, false base stations/stingrays and overshadowing as \textit{out of scope}.

\section{Methodology for Security Evaluation} \label{sec:methodology}

\begin{figure*}[!t]
     \centering
     \includegraphics[width=2\columnwidth]{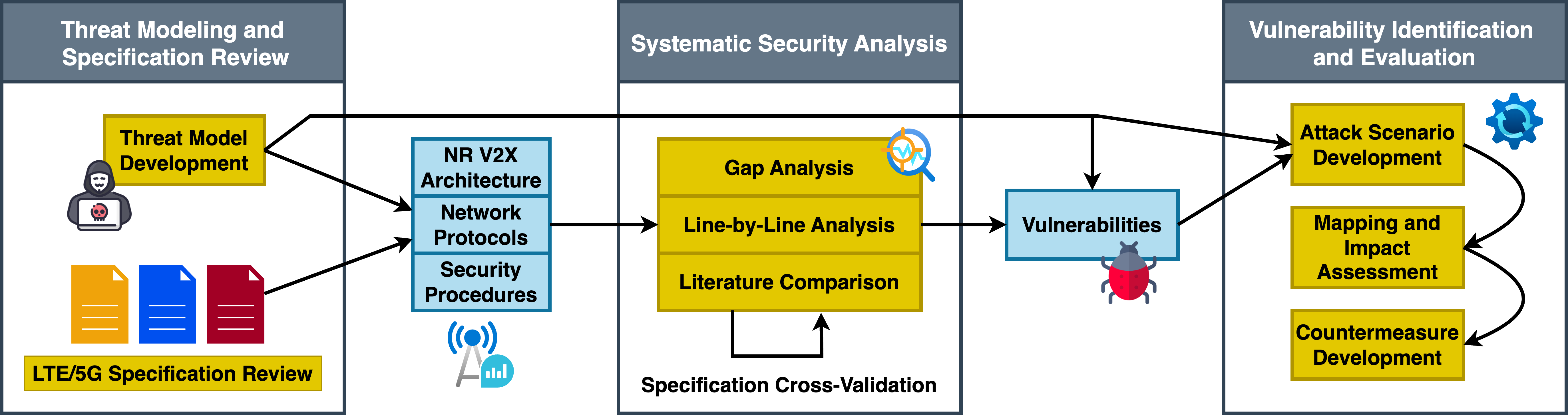}
     \caption{The figure illustrates our methodology; (1) the threat modeling and reviewing of specification documents, (2) the various security analysis techniques used, and (3) the identification and evaluation of vulnerabilities.}
     \label{fig:method}
\end{figure*}

This section outlines the systematic approach we employed in order to provide a robust framework for assessing the 5G NR V2X sidelink communication system. Figure~\ref{fig:method} provides a high-level overview of the methodology.

\subsection{Threat Modeling and Specifications Review}

\noindent\textbf{Threat Model Development.} Building on the threat model established previously, we focused on adversaries capable of intercepting, modifying, or transmitting sidelink messages. These include both compromised UEs and malicious actors with valid network subscriptions. The defined threat model highlights scenarios such as impersonation, message injection, and resource blocking, which were critical to guiding our analysis. Specifically, the threat model directed attention to areas where malicious actors could exploit synchronization signals, HARQ feedback, and resource allocation mechanisms, helping prioritize vulnerabilities with the highest potential impact. 

\noindent\textbf{Comprehensive Review of 3GPP Specifications.} We conducted an in-depth review of key 3GPP specifications, including 33.536~\cite{3gpp.33.536} (Security aspects of NR V2X services), 38.213~\cite{3gpp.38.213} (Physical layer procedures for control), and 38.331~\cite{3gpp.38.331} (RRC protocol specification). Guided by the threat model, this analysis targeted areas critical to sidelink communication, such as synchronization, resource allocation, and integrity mechanisms. This involved analyzing the defined procedures, protocols, and security mechanisms governing NR V2X sidelink communications, as well as their interplay with the conventional cellular architecture, to uncover potential vulnerabilities and ensure comprehensive security coverage. Emphasis was placed on identifying gaps where security protections were insufficient or absent, particularly in scenarios outlined by the threat model.

\subsection{Systematic Security Analysis}

\noindent\textbf{Gap Analysis.} This process involves identifying security guarantees and mechanisms that are either missing or incomplete in the official 3GPP specifications. In our work, we compared the 3GPP-defined security measures against fundamental security aspects (e.g., confidentiality, integrity, availability, authenticity, etc.), encouraged by established frameworks like STRIDE~\cite{hernan2006stride} and NIST~\cite{nist2018framework}. For instance, while 3GPP provides robust protections at higher layers, it does not mandate authentication for physical-layer messages such as Sidelink Control Information (SCI), leaving them vulnerable to spoofing and manipulation.

\noindent\textbf{Line-by-Line Analysis and Literature Comparison.} We conducted a meticulous line-by-line examination of key 3GPP specifications, including TS 33.536~\cite{3gpp.33.536} (Security aspects of NR V2X services), TS 38.213~\cite{3gpp.38.213} (Physical layer procedures for control), and TS 38.331~\cite{3gpp.38.331} (RRC protocol specification). %This detailed review allowed us to identify vulnerabilities within synchronization, resource allocation, and message integrity procedures. 
Parsing individual clauses provided granular insights into the technical implementation, helping us uncover security-sensitive areas that might otherwise be overlooked. To validate and expand upon these findings, we incorporated insights from academic and industry literature. This comparative analysis not only confirmed observations from our specification review but also identified vulnerabilities that persist from earlier systems or are amplified in NR V2X.

\noindent\textbf{Specification Cross-Validation.} We cross-referenced each layer's or procedure's documents with security documents within the 3GPP standards to verify whether each layer’s defined procedures align with or contradict higher-level security requirements. For instance, we compared references in the physical-layer (e.g., TS 38.213~\cite{3gpp.38.213}) against security specifications (e.g., TS 33.536~\cite{3gpp.33.536}) to check if security controls mandated at upper layers were actually enforced below. While TS 33.536 references key management for sidelink, we found no mention in TS 38.213 requiring authentication or integrity for the physical-layer. 

Similarly, we conducted cross-validation for all layers in NR V2X. Through this process, we pinpointed mismatches where protocol architecture fails to propagate security requirements downward/upwards, ultimately revealing the vulnerabilities discussed in the sections later.

\subsection{Vulnerability Identification and Evaluation}

\noindent\textbf{Attack Development.} Guided by our threat model, we transformed each discovered vulnerability into an attack by considering the specific resources and capabilities an adversary requires. By mapping each vulnerability to a realistic exploit pathway, we also established the goal of the attacker in each case. For example, the absence of cryptographic checks for HARQ feedback lead to HARQ spoofing.

\noindent\textbf{Mapping and Impact Assessment.} Each identified vulnerability was mapped to specific attack vectors and evaluated for its threat and impact. This included analyzing technical requirements for executing the attack and the potential consequences on network performance and critical applications.

\noindent\textbf{Countermeasure Development.} For each identified vulnerability and associated attack, we proposed targeted countermeasures. 
For example, implementing authentication and integrity protection for critical control messages, and advocating for updates to the 3GPP specifications. Potential overhead and latency implications should be considered as well.

\section{Sidelink Synchronization Attacks} \label{sec:sync}

\subsection{Synchronization Procedure}

The synchronization process~\cite{3gpp.37.985} includes primary synchronization sources such as the gNodeB or GNSS, which typically provide the timing references. However, when direct access to these sources is unavailable, SyncRef UEs step in to maintain timing coherence within the sidelink network. In this role, SyncRef UEs the S-SSBs, that convey timing references to surrounding UEs, ensuring they can align their transmission timing and frequency with each other.

\begin{figure}[!t]
     \centering
     \includegraphics[width=\columnwidth]{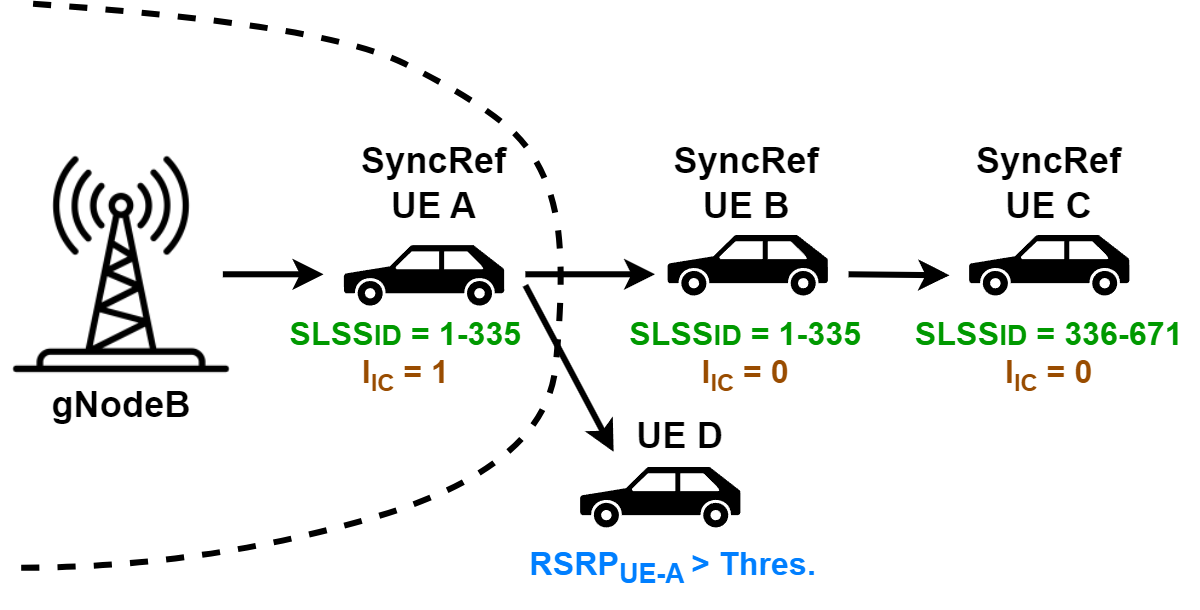}
     \caption{An example of a cellular-based sidelink synchronization stage. The $SLSS_{ID}$ is not 0 for UE A, because it synchronizes with a gNodeB, not GNSS.}
     \label{fig:sidelink-net}
\end{figure}

Figure~\ref{fig:sidelink-net} illustrates this process with multiple UEs operating within a synchronization hierarchy. Here, SyncRef UE A acts as a primary reference for synchronization and broadcasts the S-SSB with an $SLSS_{ID}$ of 1-335, marking it as in-coverage. Other UEs, such as SyncRef UE B, synchronize to UE A, adopting a $SLSS_{ID}$ within 1-335 too and further relaying the timing information to surrounding UEs. It should be clarified that an $SLSS_{ID}$ can be equal to 0, and can be used by UEs that are either directly synchronized (like UE A) or second-level synchronized (like UE B) with GNSS only. In contrast, SyncRef UE C operates out of coverage, signified by its $SLSS_{ID}$ of 336-671. The \(I_C\) value indicates the synchronization priority, with \(I_C = 1\) representing direct synchronization with the primary source (e.g., GNSS or gNodeB) and \(I_C = 0\) for UEs synchronized indirectly through another SyncRef. This parameter is included in the Master Information Block Sidelink (MIB-SL)~\cite{3gpp.38.331}, which is transmitted together with the SLSS. The MIB-SL is a crucial message that includes the system information transmitted by a SyncRef UE, Table~\ref{tab:mib-sl} denotes its contents. The above hierarchy is further reinforced by the PSBCH-RSRP (Physical Sidelink Broadcast Channel - Reference Signal Received Power), where UEs select the strongest signal that meets a threshold to ensure a reliable timing reference. Simply, the UE measures the RSRP of the Demodulation Reference Signals (DM-RS) embedded in the PSBCH.

\begin{table}[!t]
\centering
\caption{MIB-SL Fields \& Sizes}
\label{tab:mib-sl}
\renewcommand{\arraystretch}{1.2}
\small
\begin{tabular}{c c}
\toprule
\textbf{Field} & \textbf{Size (bits)} \\
\midrule
\texttt{sl-TDD-Config-r16} & 12 \\
\texttt{inCoverage-r16} & 1 \\
\texttt{directFrameNumber-r16} & 10 \\
\texttt{slotIndex-r16} & 7 \\
\texttt{reservedBits-r16} & 2 \\
\bottomrule
\end{tabular}
\end{table}

Furthermore, the $SLSS_{ID}$ is a key component within the S-SSB that uniquely identifies a SyncRef UE and conveys its synchronization priority. The $SLSS_{ID}$ is derived from a specific combination of S-PSS and S-SSS sequences, with 2 possible S-PSS sequences and 336 possible S-SSS sequences, resulting in a total of 672 unique values. This identifier also allows the receiving UEs to determine the most suitable synchronization reference and prevent conflicts based on its priority. Unlike the NR Uu interface, which uses a random access procedure to notify the gNB of the UEs' presence, NR V2X sidelink lacks such a procedure, meaning that the SyncRef UE remains unaware of which UEs have successfully synchronized with it. If no synchronization source is available, the UE defaults to using its internal clock.

The decision for a UE to become a SyncRef UE and transmit S-SSBs is based on specific RSRP measurements. There are two main procedures for initiating S-SSB transmissions. First, a UE may be explicitly configured by the network (e.g., a gNodeB) to act as a SyncRef UE. If configured, the UE will continuously transmit S-SSBs regardless of whether it has sidelink data to transmit. Alternatively, if not explicitly configured, the UE may autonomously decide to transmit S-SSBs based on the RSRP of the synchronization signals. If the RSRP is below a predefined threshold, indicating weak or no coverage, the UE transmits S-SSBs; otherwise, it refrains from doing so. The UE selects the $SLSS_{ID}$ and the slot in which to transmit the SLSS. This approach allows UEs at the edge of network coverage to become SyncRefs, extending synchronization coverage to nearby UEs that lack a direct connection to the network.

In the example, standard UEs like UE D determine synchronization by evaluating the RSRP of signals received from SyncRef UEs, choosing the one with the highest power that surpasses a predefined threshold. This selection ensures stable synchronization across UEs, even in scenarios without a direct connection to the core network. The synchronization process is therefore organized as a relay-based hierarchy, where SyncRef UEs extend network coverage by serving as timing references, enabling consistent, low-latency communication across the sidelink network. 

Ultimately, the \texttt{SL-SyncConfig Information Element}~\cite{3gpp.38.331} is important as it provides all the necessary parameters for reception and transmission of sidelink synchronization signals, and includes the sl-SSID (i.e, $SLSS_{ID}$), sl-SyncRefMinHyst (threshold for syncRef UE evaluation, as in Figure~\ref{fig:sidelink-net}), sl-SyncRefDiffHyst (threshold for SyncRef UE evaluation in reselections), and the syncTxThreshOoC (threshold for signal transmission).

\subsection{Security Issues in Sidelink Synchronization Procedures}

5G NR sidelink synchronization has security weaknesses due to unauthenticated identifiers, static configurations, and lack of control over its broadcasts, exposing it to various attack vectors.

\subsubsection{Unauthenticated Identifiers and Vulnerable Broadcasts} Key identifiers like the $SLSS_{ID}$ and $I_{IC}$, essential for SyncRef UEs, lack authentication and integrity checks. This makes it easy for malicious UEs to impersonate legitimate SyncRefs, broadcasting counterfeit synchronization signals that desynchronize legitimate UEs. S-SSB broadcasts are also unprotected, enabling attackers to inject false synchronization information, particularly impactful in out-of-coverage areas where SyncRefs are the primary timing sources. This also includes the Master Information Block for Sidelink (MIB-SL). These transmissions of S-SSBs rely on RSRP thresholds and are not fully bound to explicit network authorization. The absence of authentication in both identifiers and broadcast messages enables unauthorized devices to interfere with network timing and mislead UEs causing desynchronization and disruptions.

\subsubsection{Static Synchronization Hierarchy and Manipulable Priority} The synchronization hierarchy is static and based on SyncRefs, when primary sources are not available. The reliance on a static synchronization hierarchy, where synchronization priority is determined solely by unauthenticated identifiers such as $SLSS_{ID}$, $I_{IC}$, and $RSRP$, introduces vulnerabilities. For example, attackers can amplify their signal strength to manipulate $RSRP$ or the identifiers, tricking UEs into prioritizing their signals over legitimate sources. This rigid structure, without adaptive measures, allows attackers to exploit the system, especially in out-of-coverage or lightly monitored areas, where network-based coordination is limited.

\subsubsection{Inadequate Control Over SyncRef Roles and Authorization} The sidelink system lacks enforcement mechanisms to regulate the number and location of active SyncRef UEs, giving attackers the opportunity to deploy multiple rogue SyncRefs, increasing the risk of synchronization conflicts and interference. Additionally, out-of-coverage UEs rely on pre-configured settings and lack real-time authorization, which enables unauthorized UEs to participate in sidelink communications without verification (such as policies in the policy control function), exacerbating the risk of communication disruptions in critical applications. The specifications do not also provide mechanisms for real-time authorization or revocation in such cases, or actual control over the transmissions. While TS 33.536~\cite{3gpp.33.536} specifies procedures for authorization and provisioning of parameters, it acknowledges limitations in policy activation too, which affect out-of-coverage scenarios (pre-configured parameters) as well, such as hardware constraints [Clause 5.3.3.1.4.2.3].

\subsection{Attack: Synchronization Abuse}

\begin{figure}[!t]
     \centering
     \includegraphics[width=\columnwidth]{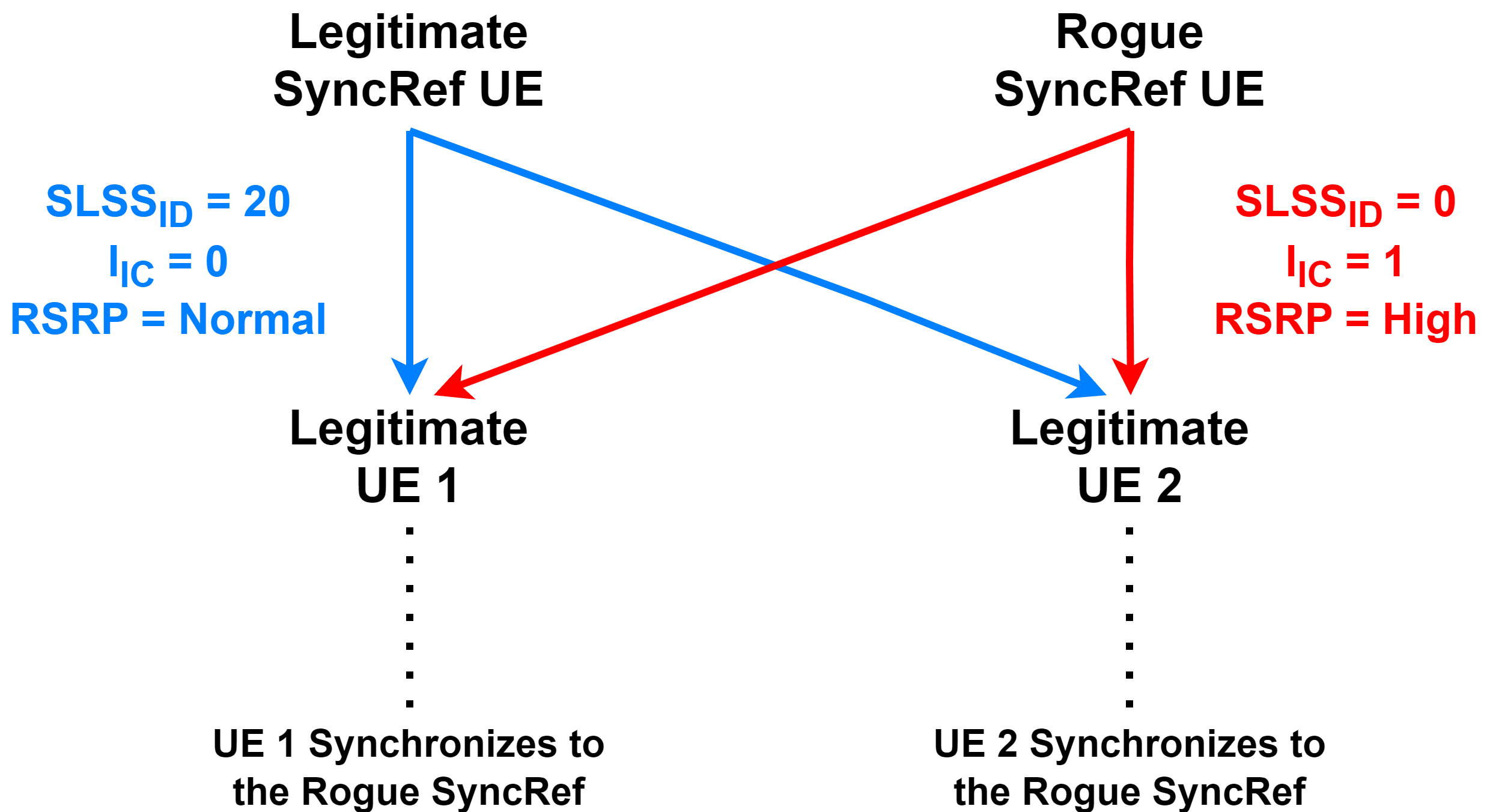}
     \caption{A simple depiction of a false synchronization reference injection.}
     \label{fig:false-injection}
\end{figure}

The 5G NR sidelink architecture is vulnerable to specific attacks due to a lack of robust authentication and integrity protections. These vulnerabilities, especially in synchronization procedures, open the network to malicious actions by unauthorized UEs. This section describes two primary attacks that illustrate how these weaknesses can be exploited to disrupt network reliability and security.

\subsubsection{Impersonation of SyncRef UE}

This attack involves a malicious UE exploiting the lack of authentication in SLSS transmission to impersonate a legitimate SyncRef UE. The attacker broadcasts SLSS messages using an arbitrary synchronization identifier, $SLSS_{ID}$, already present in the network, falsely indicating synchronization to a superior reference source. It transmits these signals at elevated power levels to ensure that its RSRP surpasses the threshold required for synchronization at nearby UEs. Due to the synchronization hierarchy, legitimate UEs will prioritize signals with higher RSRP, synchronizing their timing and frequency references to the attacker’s malicious signal instead of an authorized SyncRef UE. This desynchronization causes UEs to misalign with the legitimate network timing, resulting in increased error rates, communication failures, and potential interference in critical sidelink applications. The impact can further propagate as affected UEs that synchronize to the attacker may inadvertently become SyncRefs, extending the disruption throughout the network.

\subsubsection{False Synchronization Reference Injection}

In this case (Figure~\ref{fig:false-injection}), a malicious UE broadcasts entirely fabricated synchronization signals to act as a false high-level SyncRef. The attacker broadcasts its own fake synchronization signals with $SLSS_{\text{ID}}$ values specifically within the valid range ${0, 1, \dots, 335}$ and setting the indicator $I_{IC} = 1$, which indicates direct synchronization to a primary source. The attacker can choose the $SLSS_{\text{ID}} = 0$ to denote synchronization with GNSS. By transmitting fake SLSS and MIB-SL messages at high power, the attacker ensures that the signal's RSRP surpasses the threshold for synchronization selection among nearby UEs. Consequently, legitimate UEs in proximity to the attacker synchronize with this false timing reference, and may maliciously attach to it. Apart from critical communication errors and failures in high-stakes applications, this potential attachment may open the door for further active exploitation by the attacker. This attack can leverage the synchronization prioritization mechanism rather than merely impersonating an existing SyncRef, focusing more on attracting UEs to its signal directly.

\section{Resource Allocation Attacks} \label{sec:resource}

\subsection{Overview of Resource Allocation}

Resource allocation determines how UEs access and utilize radio resources over the PC5 interface for direct communication. 
In Mode 1 (Network-Scheduled), the gNodeB centrally assigns resources, specifying parameters like frequency bands, time slots, modulation, and power levels through SIBs or RRC signaling (e.g., \texttt{RRCReconfiguration messages}). This centralized approach enables optimized resource utilization, interference management, and coordination, which are essential in high-density scenarios. 

In Mode 2 (UE-Autonomous), UEs operate without direct network assistance, managing their own resources within pre-configured resource pools. Each UE employs sensing mechanisms to detect occupied resources, measuring energy levels or decoding Sidelink Control Information (SCI) messages from neighboring UEs. Based on this sensing data, UEs apply dynamic or Semi-Persistent Scheduling (SPS) to select unoccupied resources, minimizing collisions and interference. SCI messages are transmitted over the PSCCH (Physical Sidelink Control Channel) and indicate the frequency-time resources a UE has selected.

\subsection{Security Issues in Resource Allocation}

Since SCI messages used to announce resource reservations in NR V2X sidelink communications lack authentication and integrity protection, attackers can exploit this vulnerability by transmitting false SCI messages over the PSCCH. These SCI messages contain critical parameters for resource allocation, such as time-frequency resource assignments, Resource Reservation Interval (RRI), and Priority, which inform neighboring UEs about the resources the transmitting UE intends to use and for how long. To be more specific, the attacker should target the SCI 1-A format (see Table~\ref{tab:sci1a-format}) which includes the necessary parameters for resource reservation. The RRI is included in the Resource Reservation Period.

Consequently, by manipulating these parameters in counterfeit SCI messages, attackers can mislead legitimate UEs into believing that certain resources are reserved when they are not. This manipulation disrupts the autonomous resource selection mechanisms, particularly in Mode 2 operations where UEs rely heavily on received SCI messages and sensing for resource selection.

\subsection{Attack: Resource Blocking}

\begin{figure}[!t]
     \centering
     \includegraphics[width=\columnwidth]{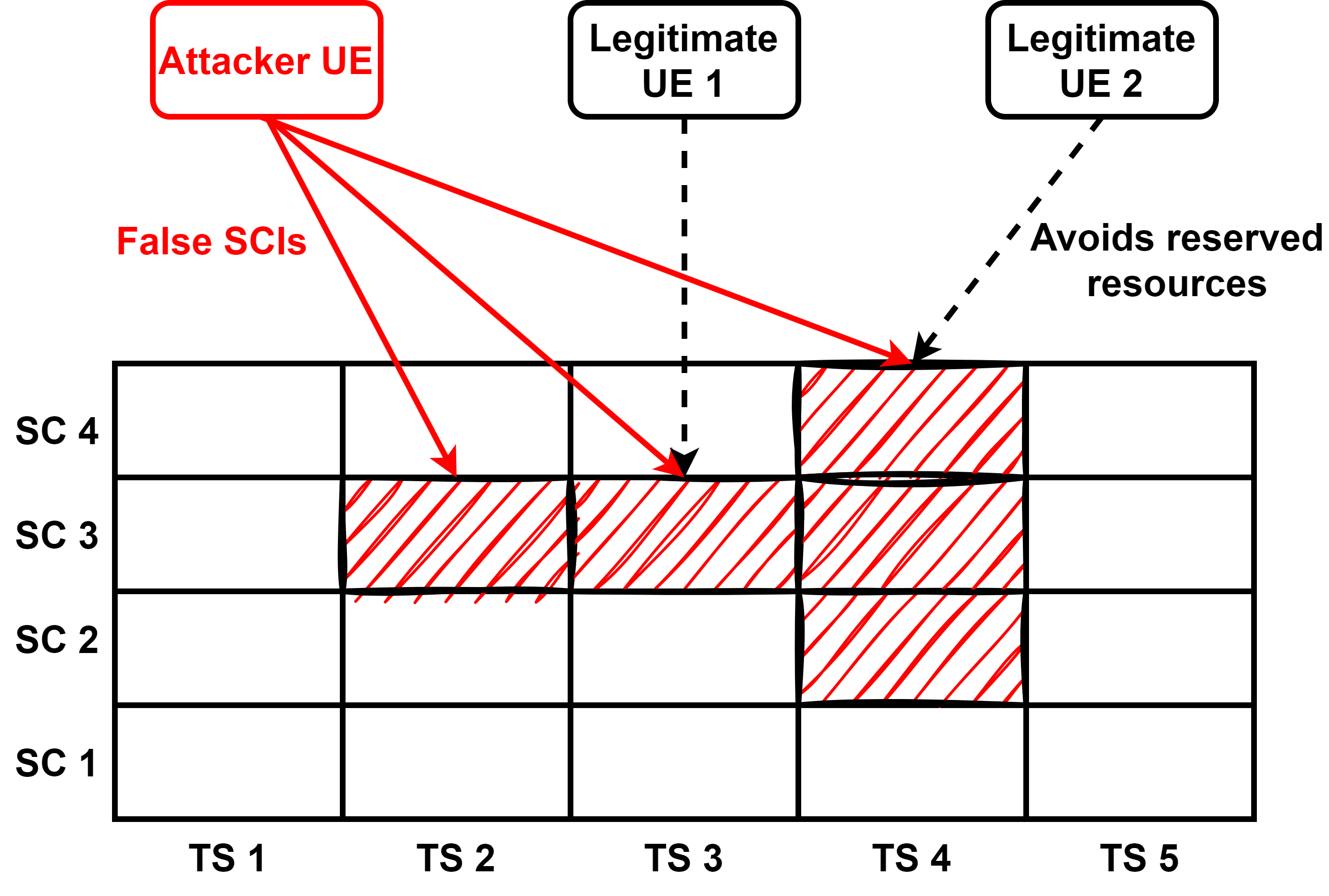}
     \caption{Resource reservation of Subchannels (SC) and Time Slots (TS) against legitimate UEs. Such attacks could cause connection disruptions or failures that may affect other vital processes.}
     \label{fig:resource-exhaustion}
\end{figure}

The primary objectives of a resource allocation attack (Figure~\ref{fig:resource-exhaustion}) are two-fold: i) \emph{Claiming Frequency Subchannels and Time Slots}: The attacker signals through SCI messages that most frequency subchannels are reserved, spanning multiple time slots, creating an artificial scarcity of available resources. ii) \emph{Extended Resource Reservation Intervals (RRIs)}: By setting RRIs to the maximum permissible duration, the attacker locks down these subchannels for prolonged periods, preventing legitimate UEs from accessing them.

The attack begins by the attacker acquiring the resource pool configuration , which specifies the frequency subchannels and time slots available for Mode 2 communication. Using spectrum sensing, the attacker detects active transmissions (e.g., related SIBs and SCIs, which are not protected), and then with knowledge of the resource pool (e.g., \texttt{SL-ResourcePool Information Element}), generates and transmits fake SCI messages to nearby UEs. By falsely claiming multiple subchannels and consecutive time slots, the attacker marks a significant portion of the spectrum as unavailable. To extend the impact, the attacker sets the RRI in the fake SCI messages to the maximum allowed (e.g., up to 1000 ms, as defined by NR V2X specifications). This extended reservation blocks subchannels for longer durations, reducing legitimate UE access. By using larger RRIs, the attacker minimizes their own transmission frequency, although periodic updates are still required to maintain the illusion of continuous occupancy.

Additionally, the attacker can monitor the radio environment to assess congestion levels and observe legitimate UE responses. By analyzing channel activity and delays, the attacker adapts their strategy to maximize congestion. As a result, legitimate UEs must select from a reduced pool of resources, increasing collision and retransmission probabilities. Although UEs employ collision avoidance, while the attacker needs precision, the artificially induced congestion can lead to transmission delays and increased message collisions. In dense environments, the attack can escalate to a DoS, severely impacting safety-critical applications reliant on sidelink communication.

Despite its impact, executing this attack requires: (1) The attacker must inject SCI messages within the correct transmission windows to ensure they are processed by legitimate UEs before resource selection occurs, (2) The adversary must craft correctly the parameters, such as the subcarrier spacing, resource pool configuration, and RRI settings for successful manipulation, (3) The attacker may need a higher transmission power, and (4) The attacker must continually inject false SCI messages, as legitimate UEs will eventually re-evaluate resources based on sensing and periodic re-selection procedures.

\section{Reliability and Integrity Attacks} \label{sec:harq}

\subsection{HARQ Feedback Procedure}

\begin{figure}[!t]
     \centering
     \includegraphics[width=0.9\columnwidth]{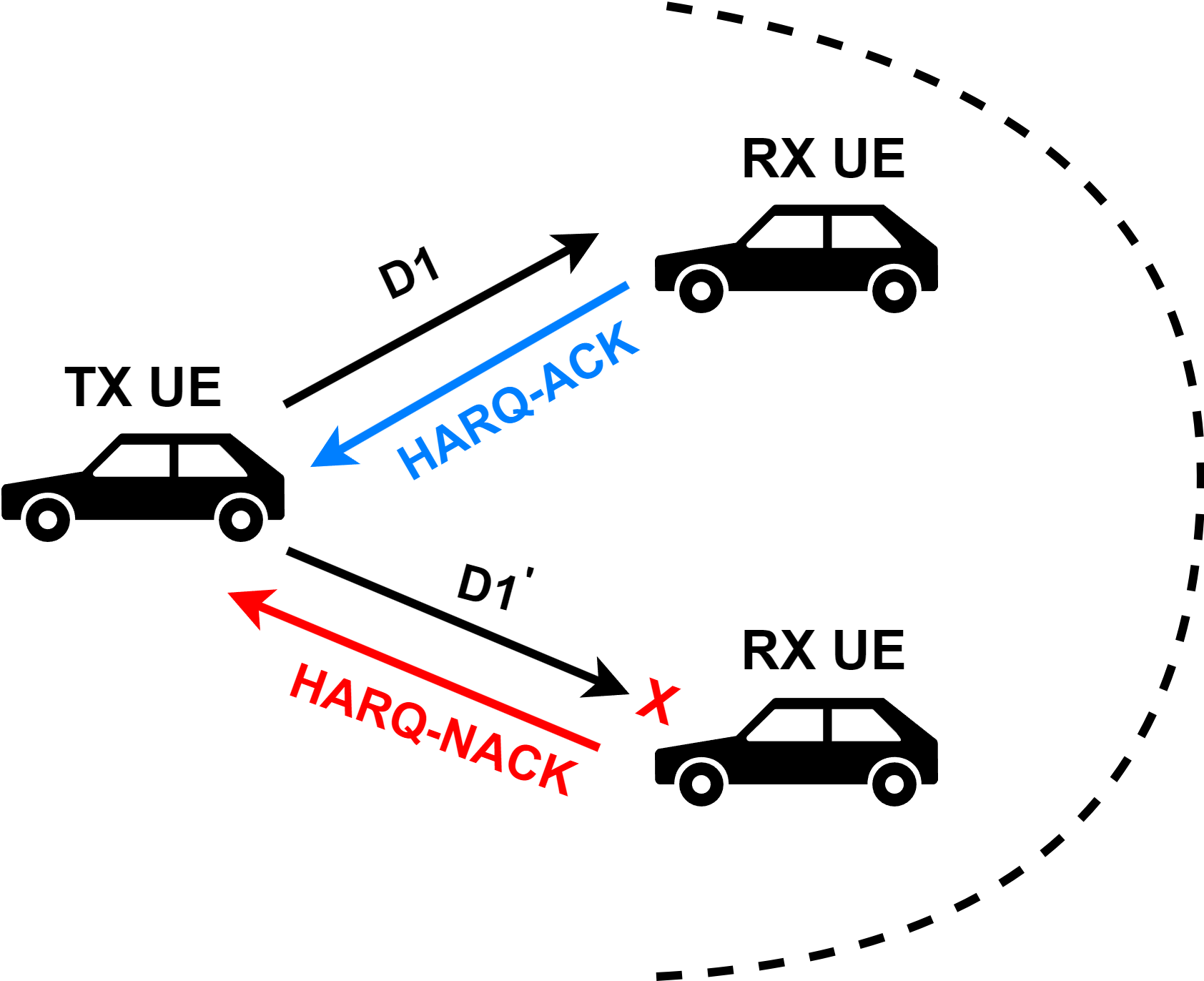}
     \caption{HARQ feedback process for NR V2X autonomous communication.}
     \label{fig:harq}
\end{figure}

5G NR sidelink introduces HARQ mechanisms for direct device-to-device communication, unlike its LTE predecessor. The HARQ process, crucial for reliable data transmission, involves the transmission of Transport Blocks (TBs) over the Physical Sidelink Shared Channel (PSSCH). Upon receiving a TB, the receiving UE verifies its integrity using a Cyclic Redundancy Check (CRC) and provides HARQ feedback via the Physical Sidelink Feedback Channel (PSFCH). This feedback can either be an acknowledgment (ACK) indicating successful reception or a negative acknowledgment (NACK) indicating a need for retransmission. HARQ feedback messages are transmitted at predefined intervals relative to the original TB transmission specified by 3GPP~\cite{3gpp.38.213, 3gpp.38.321}, allowing for efficient error correction and minimizing higher-layer protocol involvement thereby achieving ultra low latency. Figure~\ref{fig:harq} illustrates the exchanges of the HARQ messages for the transmitted data (\ie, $D_1$ and $D_{1}'$, until $D_n$ and $D_{n}'$).

\subsection{Security Issues in HARQ Process}

While HARQ is designed to enhance communication reliability, the lack of authentication and integrity protection~\cite{3gpp.37.985, 3gpp.33.536} in HARQ feedback messages introduces significant vulnerabilities. Specifically, HARQ acknowledgments (ACK/NACK) transmitted over the Physical Sidelink Feedback Channel (PSFCH) do not include any cryptographic protection, allowing attackers to forge or manipulate feedback. This absence of verification mechanisms enables malicious actors to exploit HARQ responses, forcing unnecessary retransmissions or preventing legitimate retransmissions altogether.

Additionally, SCI Format 2-A messages (format in Table~\ref{tab:sci2a-format}), which define HARQ parameters, are also unauthenticated, meaning an attacker could further modify key values such as HARQ process number, redundancy version, or new data indicator (NDI) to manipulate retransmissions, even though we consider SCI 2-A exploitation as only an enhancement to spoofing (similar to formats 2-B and 2-C~\cite{3gpp.38.212}). These weaknesses expose HARQ-based reliability mechanisms to exploitation.

\subsection{Attack: HARQ Feedback Spoofing}

In the HARQ feedback spoofing attack, an attacker exploits the unauthenticated nature of HARQ feedback. Mode 2 can be more prone to this attack due to the lack of centralized control. By injecting false NACKs, the attacker denotes that the TB was not successfully received and forces the transmitter into unnecessary retransmissions, increasing resource consumption and network latency. Alternatively, by sending false ACKs, the attacker deceives the transmitter into assuming successful data delivery (in reality, the TB was lost), causing data loss when the true receiver does not receive the TB.

Executing this attack is not infeasible but presents a few challenges: (1) the attacker must transmit the spoofed feedback exactly when the legitimate response is expected (tight HARQ window), (2) the attacker must have knowledge of the involved parameters (e.g., HARQ process number, modulation, codebook) before any attempt, and (3) the forged HARQ response must arrive at the transmitter with a stronger signal than the legitimate receiver’s feedback. Despite these obstacles, successful feedback spoofing poses significant risks, particularly in high-density scenarios where retransmission and data integrity are critical.

\section{Privacy, Authorization, and Integrity Challenges in Sidelink} \label{sec:general}

In addition to the specific synchronization, resource allocation, and feedback mechanisms vulnerabilities detailed in Sections~\ref{sec:sync},~\ref{sec:resource}, and~\ref{sec:harq}, sidelink communication in 5G NR also faces broader security and privacy challenges related to authorization, identity privacy, and message integrity. These challenges arise from inherent limitations in the PC5 interface and the flexible security policies within the NR sidelink framework, which leave communication modes --unicast, groupcast, and broadcast—susceptible to tracking, impersonation, and denial-of-service attacks. This section provides an in-depth analysis of these challenges, focusing on privacy risks associated with identity exposure, the limitations of authorization policies, and the integrity issues in PC5 messaging.

\subsection{Privacy Risks Across Communication Modes}

5G NR V2X communication supports three primary modes over the sidelink interface—unicast, groupcast, and broadcast—each serving distinct application needs but with varying privacy implications. In \emph{unicast mode}, direct communication is established between two UEs, allowing one UE to initiate a secure, private connection with a specific receiving UE. \emph{Groupcast mode} enables communication with a defined group of UEs simultaneously, essential for applications like coordinated actions or group messaging among vehicles. Finally, \emph{broadcast mode} permits a UE to transmit data to all UEs within its range, often used for safety alerts or traffic information entailing wide dissemination without specific targeting.

The privacy requirements for each communication mode vary based on the security standards set by 3GPP. For unicast mode, security requirements are stringent to ensure confidentiality, integrity, and authenticity of data between the two UEs. As specified in~\cite{3gpp.33.536} [Clause 5.3.2.1], UEs in unicast mode must establish a unique security context for each connection, creating dedicated keys and security parameters to prevent unauthorized interception and tampering. Signaling and user-plane data exchanged in unicast mode can be safeguarded with confidentiality, integrity, and replay protection, as well as measures against tracking and linkability attacks~\cite{3gpp.33.536} [Clause 5.3.2.2]. Groupcast and broadcast modes, however, have minimal security requirements. According to~\cite{3gpp.33.536} [Clause 5.4.2.1] for groupcast and [Clause 5.5.2.1] for broadcast:

\begin{fancyquote}[]
There are no requirements for securing the NR based PC5 reference point for groupcast mode.
\end{fancyquote}

\begin{fancyquote}[]
There are no requirements for securing the NR based PC5 reference point for broadcast mode.
\end{fancyquote}

For these modes, although data encryption and integrity protection are not mandated, privacy requirements remain critical due to risks of tracking. As stipulated in~\cite{3gpp.33.536} [Clause 5.4.2.2] and [Clause 5.5.2.2], UEs must guard against linkability and trackability by periodically randomizing or changing their Layer-2 IDs and IP addresses. These privacy-preserving strategies are intended to obscure UE identities, preventing long-term association or tracking of messages to the same UE.

\subsubsection{Security Flaws in Layer-2 Identifier}

Despite these privacy-preserving strategies, the frequency, level of randomization, and synchronization of identifier refreshment across layers remain undefined in the standards. This ambiguity introduces a security concern. UEs are instructed to change their Layer-2 IDs and IP addresses periodically, with randomization intended to prevent tracking or association. However, no specific guidance, method or constraints are given regarding refresh intervals or randomization methods, potentially exposing UEs to privacy risks if identifiers remain static or predictable.

To partially address this,~\cite{3gpp.23.287} suggests the use of a "privacy timer," allowing UEs to self-assign Layer-2 IDs based on a timer that specifies when an identifier change should occur:

\begin{fancyquote}[]
A privacy timer value indicating the duration after which the UE shall change each source Layer-2 ID self-assigned by the UE when privacy is required.
\end{fancyquote}

However, Layer-2 IDs in MAC layer headers are typically transmitted in plaintext, as they are required for routing at the physical layer. Since encryption occurs only at the PDCP layer, Layer-2 IDs remain visible over-the-air, increasing the potential for tracking. This issue begins as a design deficiency and propagates to the implementation side as well.

\subsubsection{Attack: UE Tracking}

The lack of secure, frequent randomization for Layer-2 identifiers introduces significant risks of UE tracking. Attackers can passively or actively monitor the PC5 interface, using software-defined radios (SDRs) or other equipment to capture Layer-2 frames. By decoding these frames, attackers can extract Layer-2 IDs and associate them with specific UE attributes, such as signal strength and transmission patterns, or even with Application-Layer IDs, creating detailed profiles and monitoring UE behaviors across locations. 

When Layer-2 identifiers are static for prolonged periods, the risk of tracking is amplified, allowing attackers to track UEs continuously. Even with identifier randomization, if the randomization process is insufficiently robust or predictable, attackers may still correlate new identifiers with old ones, effectively bypassing privacy defenses. This attack, underscores the serious privacy implications, as sustained tracking could reveal a UE's location and movement patterns, posing significant privacy risks for users.

\subsection{Service Authorization, Control and PC5 Weaknesses}

This section highlights how service provisioning and insecure settings introduce vulnerabilities, which can be exploited for attacks.

\textbf{Role of the Policy Control Function.} The PCF is critical in handling V2X service authorization, in Mode 1, where it provisions UEs with V2X policies based on their PC5 capabilities~\cite{3gpp.23.287}. It determines the default communication mode (broadcast, groupcast, or unicast) for each V2X service type, assigns authorization and policy parameters, and maps service types to frequencies and geographical areas. Furthermore, the PCF supplies Quality of Service parameters to the AMF, retrieves V2X data from the UDR to align with the subscriber’s profile, and delivers privacy policies, including Layer-2 ID mapping requirements across modes.

\textbf{Authorization and Policy Provisioning Procedures.} During the UE registration process, defined in~\cite{3gpp.23.502} [Clause 4.2.2.2], the UE indicates its V2X capability. If authorized based on subscription data, the AMF selects a PCF supporting V2X policy provisioning~\cite{3gpp.23.287} [Clause 6.2.2]. A UE policy association is then established, allowing the PCF to provision V2X policies and parameters using the Policy Association Establishment procedure~\cite{3gpp.23.502} [Clause 4.16.11]. The policies may be updated if (1) the UE switches to a new PLMN, (2) the subscription data changes, or (3) specific service parameters require modification~\cite{3gpp.23.502} [Clause 4.16.12.1].

\begin{figure}[!ht]
     \centering
     \includegraphics[width=\columnwidth]{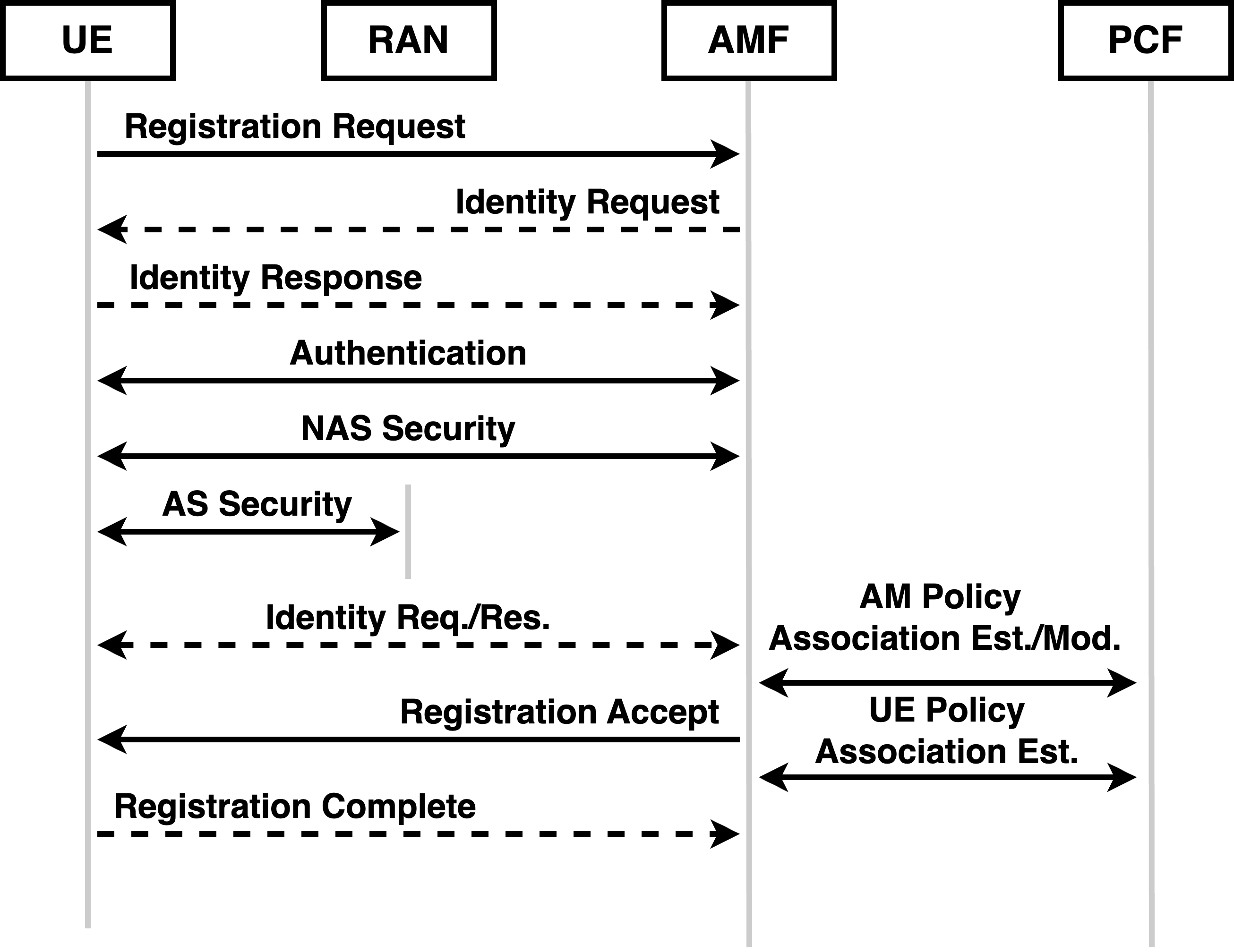}
     \caption{Registration establishment between the UE and the network. The UE and network perform the authentication and key agreement before the PCF communication.}
     \label{fig:5gs-connection}
\end{figure}

If the current V2X policies are outdated or missing, the UE initiates policy provisioning in Mode 1 after registration~\cite{3gpp.23.287} [Clause 6.2.4]. This ensures the UE operates with authorized configurations, prioritizing parameters from the PCF. If unavailable, the UE relies on parameters from the V2X Application Server, UICC, or pre-configured settings, while also adhering to regional frequency and geographical regulations. In Mode 2 (out-of-coverage), UEs rely on pre-configured parameters for V2X communication. This includes radio parameters for PC5 RAT and privacy timer values, allowing UEs to operate without direct access to a 5G Core. Figure~\ref{fig:5gs-connection} depicts the PCF participation in the UE registration.

\begin{figure}[!t]
     \centering
     \includegraphics[width=0.9\columnwidth]{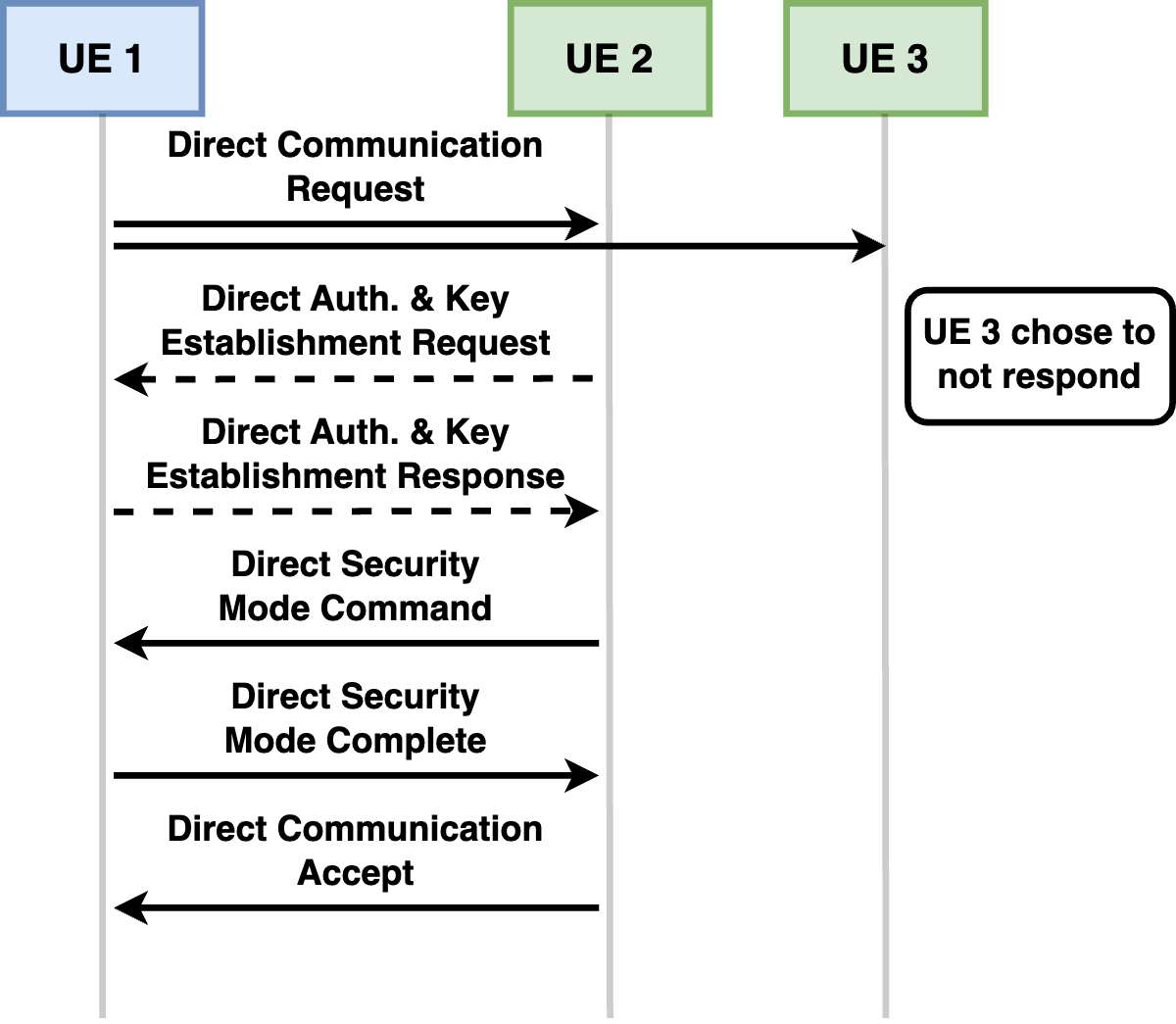}
     \caption{Connection establishment between UEs. UE-1 initiates the connection with UE-2 and UE-3, but only UE-2 establishes a full connection. UE-3 ignores UE-1's request.}
     \label{fig:connection}
\end{figure}

\textbf{Connection Establishment.} To initiate a unicast PC5 link, the UE begins by determining the destination Layer-2 ID and other necessary parameters from the V2X application layer. As outlined in~\cite{3gpp.33.536}, the initiating UE sends a \texttt{Direct Communication Request} containing Source User Info, V2X Service Info, and optional Target User Info. Authentication may occur, depending on the security policies of the UEs, using \texttt{Direct Authentication Request} and \texttt{Direct Authentication Response} messages. After mutual verification, secure link establishment follows with \texttt{Direct Security Mode Command} and \texttt{Direct Security Mode Complete}, ensuring that the communication is protected by unique ciphering and integrity keys (Section~\ref{sec:management} in the Appendix discusses key management further). Upon successful setup, the target UE sends a \texttt{Direct Communication Accept} to complete the process. UEs are permitted to ignore communication requests, as shown by UE-3 in Figure~\ref{fig:connection}, effectively rejecting the link establishment.

\textbf{Protection of PC5 Messages.} 3GPP~\cite{3gpp.24.587} [Clause 6.1.2.11.3] specifies that:

\begin{fancyquote}[]
If the signaling integrity protection is not activated for PC5 unicast link, all PC5 signaling messages are processed by the UE without integrity protection.
\end{fancyquote}

\begin{figure}[!t]
     \centering
     \includegraphics[width=\columnwidth]{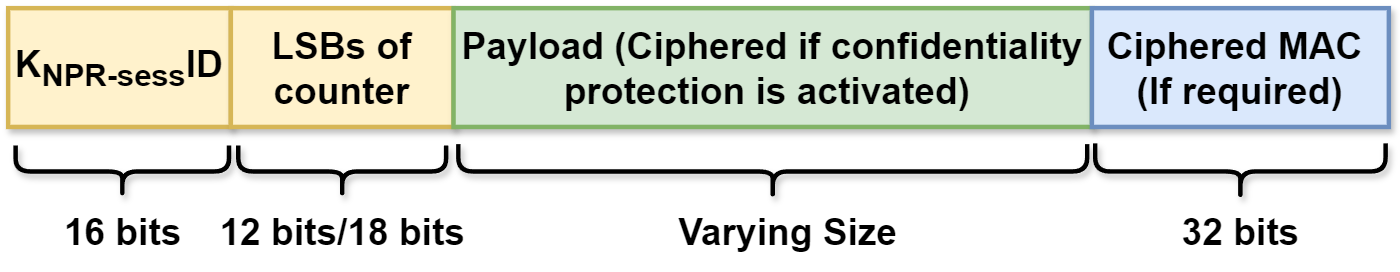}
     \caption{Parameters in the PDCP header.}
     \label{fig:pdcp}
\end{figure}

This provision applies when UEs have not yet established security methods for the PC5 interface. As a result, this also affects the RRC layer, as ciphering and integrity protection applies to the messages at the PDCP layer, as illustrated in Figure~\ref{fig:pdcp}~\cite{3gpp.33.536}. Once a secure context has been established between UEs over the PC5 interface, all subsequent signaling messages must be protected by both encryption (ciphering) and integrity checks. This ensures confidentiality and message authenticity, preventing eavesdropping, tampering, and replay attacks. Any signaling message failing these integrity checks, or lacking required encryption, is discarded by the receiving UE to maintain secure communication. However, certain messages are accepted without protection to enable UEs to negotiate and establish security mechanisms. These unprotected messages, sent before security is fully established, can pose significant vulnerabilities. Table~\ref{tab:pc5_messages} lists all PC5 signaling messages, including those allowed to be transmitted without encryption or integrity protection prior to security establishment, provided non-NULL ciphers are used.

\begin{table*}[!t]
\centering
%\small
\caption{PC5 Signaling Messages and their protection, based on 3GPP~\cite{3gpp.24.587}.}
\label{tab:pc5_messages}
%\resizebox{2\columnwidth}{!}{%
\begin{tabular}{lccccc}
\toprule
\textbf{Message} & \textbf{Ciphering} & \textbf{Integrity} & \textbf{Stage} & \textbf{Definition in} \\
\midrule
1. Direct Link Establishment Request & \ding{55} & \ding{55} & Before Security & 7.3.1 \\
2. Direct Link Establishment Accept & \ding{51} & \ding{51} & After Security & 7.3.2 \\
3. Direct Link Modification Request & \ding{51} & \ding{51} & After Security & 7.3.4 \\
4. Direct Link Modification Accept & \ding{51} & \ding{51} & After Security & 7.3.5 \\
5. Direct Link Release Request & \ding{51} & \ding{51} & After Security & 7.3.6 \\
6. Direct Link Release Accept & \ding{51} & \ding{51} & After Security & 7.3.7 \\
7. Direct Link Keepalive Request & \ding{51} & \ding{51} & After Security & 7.3.8 \\
8. Direct Link Keepalive Response & \ding{51} & \ding{51} & After Security & 7.3.9 \\
9. Direct Link Authentication Request & \ding{55} & \ding{55} & Before Security & 7.3.10 \\
10. Direct Link Authentication Response & \ding{55} & \ding{55} & Before Security & 7.3.11 \\
11. Direct Link Authentication Reject & \ding{55} & \ding{55} & Before Security & 7.3.12 \\
12. Direct Link Security Mode Command & \ding{55} & \ding{51} & During Security & 7.3.13 \\
13. Direct Link Security Mode Complete & \ding{51} & \ding{51} & During Security & 7.3.14 \\
14. Direct Link Security Mode Reject & \ding{55} & \ding{55} & During Security & 7.3.15 \\
15. Direct Link Rekeying Request & \ding{51} & \ding{51} & After Security & 7.3.16 \\
16. Direct Link Rekeying Response & \ding{51} & \ding{51} & After Security & 7.3.17 \\
17. Direct Link Identifier Update Request & \ding{51} & \ding{51} & After Security & 7.3.18 \\
18. Direct Link Identifier Update Accept & \ding{51} & \ding{51} & After Security & 7.3.19 \\
19. Direct Link Identifier Update Ack & \ding{51} & \ding{51} & After Security & 7.3.20 \\
20. Direct Link Identifier Update Reject & \ding{51} & \ding{51} & After Security & 7.3.21 \\
21. Direct Link Modification Reject & \ding{51} & \ding{51} & After Security & 7.3.22 \\
22. Direct Link Establishment Reject & \ding{55} & \ding{55} & Before Security & 7.3.23 \\
23. Direct Link Authentication Failure & \ding{55} & \ding{55} & Before Security & 7.3.24 \\
\bottomrule
\end{tabular}%
%}
\end{table*}

\subsubsection{Security Issues: Authorization Challenges, Null Ciphers and Optional Authentication}

Despite the involvement of the PCF, significant authorization challenges remain, largely due to the flexibility allowed in security policies, which can undermine communication integrity. The NR sidelink architecture permits UEs to negotiate security settings, including the use of NULL ciphers that lack encryption or integrity protection. This flexibility becomes particularly risky when UEs opt also for distinct settings designated as \textit{NOT NEEDED} or \textit{PREFERRED}, leading to severely weakened security. Even UEs with different settings—some requiring mandatory security and others not—can cause mismatches and link establishment failures. Additionally, authentication—a critical step for verifying UE identity—can be bypassed, heightening the risk of MiTM, impersonation, and other attacks due to the lack of identity verification. 

\subsubsection{Security Issue: Unprotected PC5 Messages}

We identify specific PC5 signaling messages that are permitted to be sent without protection before security establishment (Table~\ref{tab:pc5_messages}). Attackers could exploit this even if the receiving UE allows the establishment of PC5 communication links under the assumption of full protection.

In substandard scenarios where protection is disabled, attackers can exploit this to launch multiple types of attacks, including DoS, impersonation, injection, MitM, and tracking. In such cases, unprotected messages such as \texttt{Direct Link Modification}, \texttt{Direct Link Release}, and \texttt{Direct Link Identifier Update} are particularly vulnerable to manipulation. Additionally, RRC layer messages are affected too by the lack of security at the PDCP layer, making them susceptible to similar attacks, since integrity protection is not enforced. This concerns the PC5-RRC specific messages, sent between UEs: (1) \textit{MeasurementReportSidelink}, (2) \textit{RRCReconfigurationSidelink}, (3) \textit{RRCReconfigurationCompleteSidelink}, (4) \textit{RRCReconfigurationFailureSidelink}, (5) \textit{UECapabilityEnquirySidelink}, (6) \textit{UECapabilityInformationSidelink}, and (7) \textit{CapabilityRequestFilterSidelink}.

\subsubsection{Attack: Exploitation of PC5 messages}

Assuming that full security has been established with mandatory authentication (otherwise, all PC5 messages are affected), the following PC5 messages can still be abused in various ways:

\begin{itemize}
    \item \textit{Impersonation via Direct Link Establishment Request}. An attacker sends forged \texttt{Direct Link Establishment Request} messages, impersonating a legitimate UE. This impersonation may cause resource exhaustion through flooding, enable MitM attacks, or lead to service disruption.
    
    \item \textit{DoS via Direct Link Establishment Reject}. Attackers can forge \texttt{Direct Link Establishment Reject} messages to block legitimate UEs from establishing connections, disrupting V2X services and leading to delays.

    \item \textit{Authentication Disruption}. By intercepting or forging \texttt{Direct Link Authentication} messages (Request, Response, Reject, or Failure), attackers can cause authentication failures or force UEs into less secure modes, exposing them to subsequent attacks.

    \item \textit{MitM During Link Establishment}. In certain conditions, attackers can intercept and modify link establishment messages, positioning themselves as a relay between UEs. This may enable eavesdropping and data manipulation, depending on the level of established security.

    \item \textit{Replay Attacks}. Attackers replay previously captured, unprotected messages, such as \texttt{Direct Link Establishment Request}, to disrupt communication or enable unauthorized actions. This is possible due to the lack of freshness checks in the initial, unprotected messages.

    \item \textit{False Security Mode Reject}. Attackers send forged \texttt{Direct Link Security Mode Reject} messages to disrupt the security establishment process. This may force UEs to abandon connections and potentially revert to less secure configurations, weakening the overall security and making subsequent messages more vulnerable.
\end{itemize}

\section{Countermeasures}

The following countermeasures aim to mitigate the vulnerabilities that have been discussed through a combination of technical measures, protocol enhancements, and robust security practices. Unfortunately, 3GPP's study on security considerations~\cite{3gpp.33.809} \textit{is currently insufficient to address these issues}, as it primarily focuses on the false base stations and conventional architecture. Key differences in the threat model—including unique UE trust to broadcast, insider threats, malicious sidelink UEs, and the use of non-smartphone devices—along with varying impacts, use cases, and network architecture, necessitate a reevaluation of both countermeasures and the broader cellular ecosystem. Section~\ref{sec:assess} of the Appendix provides more information on the feasibility and performance of each proposed measure, while focusing more on their potential challenges.

\textbf{Synchronization Protection.} To address synchronization vulnerabilities, lightweight authentication mechanisms should be implemented at the physical layer for synchronization signals transmitted by SyncRef UEs. Cryptographic signatures or certificate-based authentication embedded within these signals can ensure their authenticity and prevent attackers from mimicking SyncRef UEs or injecting false synchronization references. However, the schemes must be fully accommodated by UEs, which could be performance hurdle with design constraints.

\textbf{HARQ Protection.} HARQ feedback protection can be strengthened by validating physical-layer attributes (e.g., transmission power, modulation patterns) and setting strict timing thresholds for response validation. Assigning unique scrambling codes to each UE and employing directional antennas further mitigate feedback spoofing risks by verifying feedback origin and timing.

\textbf{Securing PC5 Messages.} Full protection for pre-security establishment PC5 messages may be impractical; instead, robust detection mechanisms should be used. Techniques like timestamps, stateful connection checks, and traffic pattern analysis detect replayed or conflicting messages. UEs should verify each connection request, as shown in Figure~\ref{fig:connection}, reducing risks of impersonation, DoS, and replay attacks. These safeguards, combined with broadcast protection, mitigate risks of malicious attachment and MitM.

\textbf{Preventing UE Tracking.} To counter identifier-based tracking, UEs should change identifiers frequently (\eg, by executing the Direct Link Identifier Update procedure) using cryptographically secure randomization and fast privacy timers. Group identifiers can help mask individual UEs within certain applications. However, performance implications of frequent identifier changes must be considered to balance security and efficiency.

\textbf{General Protocol Enhancements.} Finally, additional measures to strengthen overall security such as enforcing strict network policies, disabling NULL ciphers (except for emergency scenarios), and making authentication mandatory for communication establishment over PC5 can minimize vulnerabilities caused by flexible security preferences. Finally, UEs can record logs and maintain procedures for responding to detected security incidents, including isolation of affected UEs and notification of relevant parties.

\section{Discussion} \label{sec:discussion} Below we discuss several important (also some out-of-scope) security aspects that are relevant to NR V2X.

\textbf{Attack Requirements in Real-Life.} The exposure of cellular configurations through MIB-SL, SIBs, and other sidelink procedures enables adversaries to passively monitor and analyze the cellular environment. Specifically, attackers can infer network parameters such as resource pools, sidelink bandwidth parts, and allocation strategies. While we have emphasized the importance of integrity protection at the physical layer, we highlight that the lack of confidentiality also indirectly aids spoofing attempts—as adversaries can possess/predict system parameters, making message fabrication more effective.

We believe that our identified attacks are feasible in real-world deployments, though their execution varies in complexity. Throughout our analysis, we have outlined attack-specific constraints, noting that HARQ and SCI spoofing require stricter timing precision and synchronization, making them more complex than Synch-based attacks or PC5 message exploitation. An adversary must employ advanced radio equipment capable of rapidly decoding sidelink frames and injecting forged signals within millisecond-level timing constraints. Additionally, sufficient transmission power is often required to override legitimate signals, particularly in HARQ or SCI spoofing scenarios where the attacker competes with legitimate sidelink feedback. Physical proximity—typically within a few hundred meters—improves interception and injection capabilities, but actual attack feasibility depends on propagation conditions, hardware capabilities, and interference levels. For example, Layer-2 ID tracking attacks require an attacker to passively monitor MAC-layer headers over extended periods to correlate UE transmissions—a task that becomes more challenging in high-mobility vehicular scenarios, where signal fluctuations, and changing propagation conditions may complicate long-term tracking.

As a result, the attacker must possess a deep understanding of NR V2X parameters and the capabilities to generate valid yet deceptive messages. This makes casual eavesdroppers significantly less likely to execute sophisticated attacks, whereas well-equipped adversaries (e.g., nation states) with advanced radio hardware can effectively exploit such vulnerabilities. 

\textbf{Impact.} The identified attacks (summarized in Table~\ref{tab:mapping}) present serious risks to critical V2X applications, such as collision avoidance systems, cooperative adaptive cruise control, and emergency vehicle notifications. For instance, impersonation of synchronization reference UEs may lead to desynchronization among vehicles or drones, resulting in communication failures or delays in safety-critical messages. False synchronization references and HARQ feedback spoofing can introduce latency and reduce message reliability, jeopardizing applications that rely on real-time responsiveness. Attacks exploiting PC5 messages and Layer-2 tracking also compromise the confidentiality and authenticity of V2X communications, potentially causing resource exhaustion, data manipulation, and privacy breaches. Overall, such vulnerabilities in V2X could lead to physical harm, property damage, and accidents. Mode 2 operates without network intervention, leaving the responsibility of critical operations like authentication, integrity protection, and encryption to UEs. Consequently, this autonomy may increase susceptibility to the aforementioned various attacks.

\begin{table*}[!t]
\centering
\captionsetup{justification=centering, font=small, labelfont=bf}
\caption{Summary of the discovered vulnerabilities and attacks in 5G NR V2X.} 
\label{tab:mapping}
\renewcommand{\arraystretch}{1.0}
\small
\begin{tabularx}{\textwidth}{ 
  >{\centering\arraybackslash}m{3cm}  % Attack Column
  >{\centering\arraybackslash}m{5cm}  % Associated Vulnerabilities
  >{\centering\arraybackslash}m{3cm}  % Main Layer(s)
  >{\centering\arraybackslash}m{3cm}  % Category
}
\toprule
\textbf{Attack} & \textbf{Associated Vulnerabilities} & \textbf{Main Layer(s)} & \textbf{Category} \\ 
\midrule

\textbf{Synchronization Abuse} & 
\makecell{Unauthenticated Identifiers \\ 
Static Synchronization Hierarchy \\ 
Manipulable Priority \\
Vulnerable Broadcasts \\
Inadequate Authorization} & 
Physical & 
Design \\
\midrule

\textbf{Resource Blocking} & 
Unauthenticated SCI Messages & 
Physical / Data Link & 
Design \\
\midrule

\textbf{HARQ Feedback Spoofing} & 
Unprotected HARQ Messages & 
Physical / Data Link & 
Design \\
\midrule

\textbf{UE Tracking via Layer-2 ID} & 
\makecell{Undefined Randomization Process\\ 
Undefined Refreshment Process \\
Exposure of Layer-2 IDs} & 
Data Link / Network & 
Design \& Implementation \\
\midrule

\textbf{Exploitation of PC5 Messages} & 
\makecell{Inadequate Authorization \\ 
Null Cipher Support \\
Optional Authentication \\ 
Unprotected PC5 Messages} & 
Network & 
Design \& Implementation \\

\bottomrule
\end{tabularx}
\end{table*}

\textbf{Insider threats and compromised UEs.} NR V2X sidelink communication faces a unique risk from compromised or malicious insider UEs. Since UEs possess cryptographic keys for secure PC5 communication, a compromised UE could exploit this trust relationship to undermine network security. Current 3GPP specifications lack detailed procedures for revocation, key updates, and anomaly detection tailored to sidelink communication, leaving the network vulnerable to insider attacks. A rogue UE can misuse protected PC5 messages to degrade network integrity (e.g., imagine a UE network with 5 devices in industrial setups), as there is no mechanism to detect or mitigate these insider threats effectively. Unlike traditional cellular networks, which assume that authenticated entities are trustworthy, NR V2X networks must consider that UEs could serve both as trusted nodes and potential attackers. UEs can play dual roles—as both initiators and receivers of connections. The direct UE-to-UE communication model in sidelink exacerbates this risk, as compromised UEs can interact directly with others without passing through intermediary infrastructure. Addressing these risks necessitates robust mechanisms to detect and mitigate malicious activities originating from inside the network.

\textbf{False base stations and GNSS attacks.} 
Even in sidelink communications, false base station (e.g.,~\cite{bitsikas21handovers}) and GNSS attacks (e.g.,~\cite{Zidan21:GNSS}) remain relevant due to their role as primary synchronization sources in Mode 1. Attackers could use false base stations to broadcast incorrect configurations via System Information Block (SIB) messages types 13 (corresponds to SIB 21) and 14 (corresponds to SIB 26), for NR V2X sidelink~\cite{3gpp.38.331}, disrupting UE synchronization. For instance, these attacks may allow adversaries to associate sensitive identifiers (SUCIs/IMSI/C-RNTIs) with Layer-2 IDs for tracking. Additionally, GNSS spoofing and jamming could mislead UEs about timing and positioning.

\textbf{Bidding down attacks.} Bidding down and downgrade attacks, common in conventional cellular networks~\cite{Karakoc23:Bidding}, aim to weaken security by forcing the use of weaker ciphers (e.g., GSM/2G). However, this attack is not applicable to NR V2X sidelink due to: \textbf{(1)} Only LTE and 5G support UE-to-UE communication for proximity services, \textbf{(2)} There is no fallback mechanism over the PC5 interface, preventing interoperability with older generations/networks, and \textbf{(3)} The protocols and mechanisms differ between 5G and LTE V2X (e.g., HARQ), limiting compatibility across generations.

\textbf{Future Work.} Our investigation reveals that there is no reliable/robust and realistic (not custom) NR V2X sidelink implementation at the moment, and that the requisite testing tools (e.g., modifiable stack) are likewise unavailable~\footnote{srsRAN currently provides very limited functionalities (only LTE signal reception)~\cite{srsran}}. Despite ongoing standardization, the technology remains less widespread than advertised, compared to conventional LTE/5G implementations. Consequently, our future work will involve developing comprehensive security test frameworks for NR V2X sidelink, where we plan to implement and evaluate the proposed countermeasures as well against key metrics such as latency, throughput, and reliability. Ultimately, once commercial vehicles with NR V2X are accessible, we aim to conduct field tests in real vehicular environments.

\section{Related Work} \label{sec:related-work}

The design and implementation of LTE-V2X have been widely studied, providing an overview of long-term evolution-vehicle (LTE-V) communication and its benefits for vehicular applications~\cite{Molina17LTEVFS, bazzi21V2X, liu15cellular, Chen17v2x, Asadi14survey}. Further research has focused on the simulation environments for both LTE~\cite{Virdis16simu} and 5G networks~\cite{Liu22nr}. The architecture and capabilities of NR have been extensively explored~\cite{Garcia21tutorial}, with several studies analyzing NR performance and design implications for V2X communications~\cite{Bagheri21nr, Todisco21PerformanceAO, Tabassum23nr, Ganesan20v2x, Liu22nr}. Additionally, NR has been proposed for public safety applications~\cite{Chukhno23nr}, for military communications~\cite{Bajracharya23:military} and in support of drone operations~\cite{Mishra22CooperativeCU}, demonstrating its versatility across multiple domains, while also a few features have been implemented based on a custom Open Air Interface~\cite{elkadi23opensource5g}~\cite{oai}.

Multiple works have covered high-level overview of the security challenges~\cite{Lu20secoverview, Huang20advances, Luo20phy, alnasser19challengessolutions, marojevic18requirementsprocedures, Mohan22threats, Boualouache23challenges, GHOSAL20:V2X-Survey, Lai20:Challenges}. However, some of these studies primarily focus on "conventional" LTE/5G networks and do not dive into NR sidelink specific internal functions, while others focus on generic device-to-device communications. A survey by Yoshizawa et al.~\cite{Yoshizawa23v2xsurvey} provide a valuable overview into V2X, though it does not investigate technically the NR ecosystem, while Ying et al.~\cite{Ying24review} offer an updated literature overview. \cite{Sedar23survey, Hasan20v2x} give more insights regarding the general security posture of the V2X networks.

Various related cryptographic mechanisms and their performance in LTE and 5G communications have also been analyzed~\cite{Pizzi21securedelivery, Zhang15SeDS, Suraci21d2d, Ahmed18secure, Liu21plattoning, Alnasser20trust}, even though trust and protection is not examined holistically and at a macroscopic level, nor compatibility with 3GPP standards. Device-to-device secrecy improvements with radio resource and power management have been studied~\cite{Yiliang20secrecy}, and DoS attacks have been mathematically simulated against C-V2X resources~\cite{Trkulja20DenialofServiceAO}. Finally, Twardokus et al.~\cite{Twardokus22dos, Twardokus23dos} have notably explored resource exhaustion and jamming techniques for C-V2X targeting the resource scheduling leading to DoS, while also proposing countermeasures. On the contrary, our focus is mainly on SCI spoofing attacks (not jamming) for resources, while also proving a detailed protocol, message and parameter analysis (i.e., a unique and holistic approach) for this attack, specifically for 5G V2X Sidelink (not LTE mode 4-oriented).

Generally, our work offers the first in-depth examination of technical NR V2X procedures and protocols by focusing on their unique security implications of cellular V2X. As shown in Table~\ref{tab:related-comparison}, existing studies either address broader V2X concepts or rely on simulation/mathematical setups without fully exploring the specific functionalities.

\section{Conclusion} \label{sec:conclusion}

In this paper, we conducted a studious examination of the 3GPP specifications, providing an overview of critical physical-layer and security procedures in NR V2X sidelink communication. We identified sensitive areas requiring attention, and associated several potential attacks with them. We then assessed the impact of these attacks and proposed mitigation strategies. Our findings have been responsibly reported to GSMA and validated accordingly. This work underscores the need for security reevaluation in NR V2X and provides a foundation for future research.

\bibliographystyle{plain}
\bibliography{main}

\appendix

\section{Key Management} \label{sec:management}

Key management in NR V2X communications over the PC5 interface is essential for establishing secure unicast links between UEs. According to 3GPP~\cite{3gpp.33.536}, each UE possesses long-term credentials ($K_{long-term}$), such as symmetric keys or asymmetric key pairs, which serve as the root of trust for mutual authentication. Through mutual authentication procedures leveraging these long-term credentials, UEs derive a shared NR PC5 root key ($K_{NRP}$), forming the basis for subsequent key derivations. A 32-bit identifier known as the $KNRP_{ID}$ is associated with $K_{NRP}$ to uniquely identify the root key in communications between the pair of UEs.

From the $K_{NRP}$, UEs derive a session-specific key ($K_{NRP-sess}$) for each unicast link to ensure key freshness and session uniqueness. The derivation of $K_{NRP-sess}$ involves the exchange of nonces between the UEs—each UE generates a random nonce, and these nonces are combined during the key derivation process to introduce randomness and prevent replay attacks. A 16-bit $KNRP{sess}$ ID, constructed by combining bits selected by each UE, uniquely identifies the session key. Using $K_{NRP-sess}$, UEs derive the NR PC5 Encryption Key (NRPEK) and the NR PC5 Integrity Key (NRPIK) by applying standardized key derivation functions. These keys provide the necessary cryptographic material for confidentiality and integrity protection of both signaling and user plane data over the PC5 interface. According to [Clause 5.3.3.1.2.1]: 

\begin{fancyquote}[]
The NR PC5 Encryption Key (NRPEK) and NR PC5 Integrity Key (NRPIK) are used in the chosen confidentiality and integrity algorithms respectively for protecting PC5-S signalling, PC5 RRC signalling, and PC5 user plane data. They are derived from $K_{NRP-sess}$ and are refreshed automatically every time $K_{NRP-sess}$ is changed. 
\end{fancyquote}

\begin{figure}[!t]
     \centering
     \includegraphics[width=0.8\columnwidth]{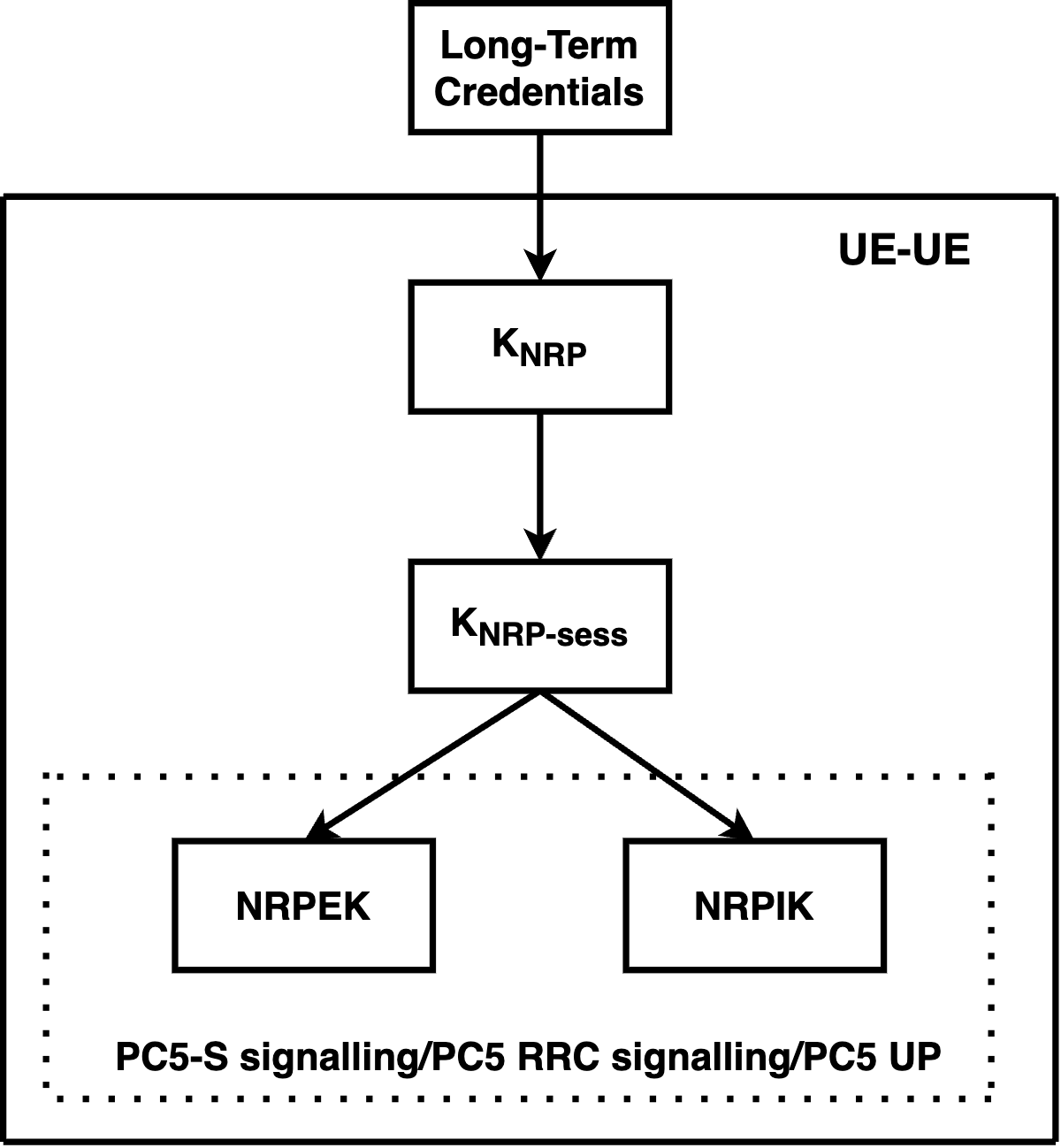}
     \caption{Key hierarchy for 5G V2X.}
     \label{fig:keys}
\end{figure}

Security contexts are established and maintained for each unicast link, encompassing the derived keys, selected algorithms, and replay protection parameters. UEs manage these security contexts throughout the communication session, updating them during rekeying procedures and securely deleting them upon session termination to prevent residual vulnerabilities. Rekeying procedures can be initiated by either UE and involve generating new nonces to derive a fresh $K_{NRP-sess}$. Additionally, identity privacy is preserved through procedures that allow UEs to change and randomize their Layer-2 IDs and KNRP IDs during active sessions, preventing tracking and linkability attacks. Figure~\ref{fig:keys} show the hierarchy of all the keys used for communication establishment based on the specifications.

\begin{table*}[t]
\centering
\caption{Comparison of Related Works on V2X Sidelink Security, Grouped by Approach}
\label{tab:related-comparison}
\renewcommand{\arraystretch}{1.2}
\small

\begin{tabularx}{\textwidth}{l l c c c c c}
\toprule
\multirow{2}{*}{\textbf{References}} & 
\multirow{2}{*}{\textbf{Approach}} &
\multicolumn{5}{c}{\textbf{NR Vulnerabilities Addressed}} \\
\cmidrule(lr){3-7}
& & 
\makecell{\textbf{Sync} \\ \textbf{Attacks}} &
\makecell{\textbf{Resource} \\ \textbf{Spoofing}} &
\makecell{\textbf{HARQ} \\ \textbf{Spoofing}} &
\makecell{\textbf{L2 ID} \\ \textbf{Exposure}} &
\makecell{\textbf{PC5} \\ \textbf{Exploits}} \\
\midrule

\multicolumn{7}{l}{\textbf{- Survey-Based Approaches (Overviews of V2X Security)}} \\
%\midrule
\makecell[l]{~\cite{Lu20secoverview, Huang20advances, Luo20phy, alnasser19challengessolutions, marojevic18requirementsprocedures, Mohan22threats, Boualouache23challenges} \\ ~\cite{Yoshizawa23v2xsurvey, Ying24review, Sedar23survey, Hasan20v2x, GHOSAL20:V2X-Survey, Lai20:Challenges}} &
\makecell[l]{Surveys, \\ Literature Reviews, \\ General Risks, Threats \\ \& Requirements.} &
$\bm{\times}$ & 
$\bm{\times}$ & 
$\bm{\times}$ & 
$\bm{\times}$ & 
$\bm{\times}$ \\

\midrule

\multicolumn{7}{l}{\textbf{- Cryptographic / Trust-Based Approaches}} \\
%\midrule
\makecell[l]{~\cite{Pizzi21securedelivery, Zhang15SeDS, Suraci21d2d, Ahmed18secure, Liu21plattoning, Alnasser20trust}} &
\makecell[l]{Key Exchange, \\ Trust Models, \\
Secure Data. }&
$\bm{\times}$ & 
$\bm{\times}$ & 
$\bm{\times}$ & 
$\bm{\times}$ & 
$\bm{\times}$ \\

\midrule

\multicolumn{7}{l}{\textbf{- Simulation-Based Approaches}} \\
%\midrule
\makecell[l]{~\cite{Yiliang20secrecy, Trkulja20DenialofServiceAO}} &
\makecell[l]{Math Analysis of DoS \\ on Resources. \\ V2V Secrecy Rate Maximization.}&
$\bm{\times}$ & 
$\bm{\times}$ & 
$\bm{\times}$ & 
$\bm{\times}$ & 
$\bm{\times}$ \\

\midrule

\multicolumn{7}{l}{\textbf{- Partially Experiment-Based Approaches}} \\
%\midrule
\makecell[l]{~\cite{Twardokus22dos, Twardokus23dos}} &
\makecell[l]{DoS Attacks against Scheduling, \\ via Jamming and Exhaustion. \\ Detection Mechanisms.} &
$\bm{\times}$ & 
$\bm{\times}$ & 
$\bm{\times}$ & 
$\bm{\times}$ & 
$\bm{\times}$ \\

\midrule

\multicolumn{7}{l}{\textbf{- Spec-Based Approaches (Detailed 3GPP NR V2X Analysis)}} \\
%\midrule
\textit{Our Work} &
Procedure \& Protocol Assessment &
\checkonly & 
\checkonly & 
\checkonly & 
\checkonly & 
\checkonly \\

\bottomrule
\end{tabularx}
\end{table*}

\section{Further Discussion on Countermeasures} \label{sec:assess}

In this section we continue the discussion about the potential countermeasures on 5G NR V2X.

\textbf{Synchronization Protection.} Authentication at the physical layer can help prevent false SyncRef signals. For instance, including cryptographic signatures or certificate-based tokens in the synchronization signals (e.g., SLSS) ensures authenticity. However, integrating such security at the physical layer introduces overhead in terms of extra bits for signatures and potential timing delays in verifying them. UEs may also need more powerful hardware or firmware support, impacting cost and battery life. This modification will require change in the design and implementation of UE stacks. An alternative is a partial integrity tag that is smaller than a full signature but still provides basic tampering detection.

Regardless, achieving this, requires embedding cryptographic material—such as a message authentication code or a short digital signature—within an extremely limited payload (i.e., signal format). According to design constraints, physical-layer sidelink messages typically have small bit budgets (e.g., tens of bits). SCIs could be more flexible in sidelink compared to the DCIs in conventional architectures (There is no DCI in sidelink, unless a gNodeB is involved in the Uu interface.), however the space limits are equally relevant. From a performance standpoint, adding any cryptographic field at the physical layer increases both computational and timing overhead. Even a small MAC calculation typically may require an extra hashing pass (e.g., HMAC with a 128-bit secret key), which must be computed by the SyncRef UE and verified by all receiving UEs within the tight synchronization window. Based on LTE/NR reference timescales, this check must be completed within milliseconds—any cryptographic validation that overshoots that boundary risks delaying the entire sidelink synchronization procedure. In resource-constrained UEs (especially in high-speed vehicular scenarios), these additional cycles could marginally raise the UE’s power consumption or reduce the effective throughput on other parallel sidelink channels. Design trade-offs need careful field evaluation to confirm that the overhead—both in bits and processing time—remains acceptable while still significantly reducing the risk of malicious synchronization injection. 

Protecting the physical layer has been discussed by past works~\cite{oh24:enablingphysicallocalization, Ross24:Broadcasts, ludant24unprotected4g5g} on conventional LTE/5G implementation, however it remains unclear whether such an implementation is applicable to sidelink and whether its adoption will be accepted. As we have already mentioned, sidelink introduces new uses cases, threats model and risks, consequently a thorough investigation of physical layer protection specifically on sidelink is paramount. Therefore, the exploration of physical layer security remains a future work.

\begin{table*}[!t]
\centering
\caption{SCI format 1-A fields for NR V2X and potential manipulations}
\label{tab:sci1a-format}
\renewcommand{\arraystretch}{1.3}
\small
\begin{tabularx}{\textwidth}{l l X}
\toprule
\textbf{Parameter} & \textbf{Bit Length} & \textbf{Attacker Manipulation \& Relevance} \\
\midrule

\textbf{Priority} &
3 bits &
Spoofing a higher/lower priority could mislead UEs about traffic importance. \\

\midrule

\textbf{Frequency Resource Assignment} &
\makecell[l]{relies on \texttt{sl-MaxNumPerReserve}} &
High impact: forging frequency allocations can cause UEs to perceive subchannels as occupied, leading to resource blocking or collisions. \\

\midrule

\textbf{Time Resource Assignment} &
\makecell[l]{5 or 9 bits \\ (relies on \texttt{sl-MaxNumPerReserve})} &
High impact: specifying multiple or extended time slots artificially reduces available resources for legitimate UEs. \\

\midrule

\textbf{Resource Reservation Period} &
\makecell[l]{$\log_2(\texttt{\#PeriodListEntries})$ bits \\ (if used)} &
High impact: setting a large reservation period (RRI) makes UEs believe resources remain taken for a long duration. \\

\midrule

\textbf{DMRS Pattern} &
$\log_2(N_{\text{DMRSPattern}})$ bits &
May affect demodulation reference signals; not crucial for blocking. \\

\midrule

\textbf{2nd-Stage SCI Format} &
2 bits &
This points to second-stage parameters, and could be used if an attacker wants to go further than just spoofing. \\

\midrule

\textbf{Beta\_offset Indicator} &
2 bits &
Modifies power offset for second-stage SCI; minimal effect on resource blocking. \\

\midrule

\textbf{Number of DMRS Port} &
1 bit &
Indicates rank-1 or rank-2 DMRS usage, not key for blocking. \\

\midrule

\textbf{Modulation and Coding Scheme (MCS)} &
5 bits &
Misrepresenting MCS might cause decoding issues, but doesn’t fundamentally block resources. \\

\midrule

\textbf{Additional MCS Table Indicator} &
\makecell[l]{1 bit (if one table) \\2 bits (if two tables) \\0 otherwise} &
References advanced MCS tables; not central for resource blocking. \\

\midrule

\textbf{PSFCH Overhead Indication} &
\makecell[l]{1 bit (if \texttt{sl-PSFCH-Period} = 2 or 4), \\ else 0} &
Might claim overhead is large, but frequency/time fields remain the main vector for blocking. \\

\bottomrule
\end{tabularx}
\end{table*}

\textbf{HARQ Protection.} Similarly, protecting HARQ feedback (ACK/NACK) with cryptographic material from spoofing entails adding integrity checks or authentication tokens to a message that is notoriously small and time-sensitive. Typical HARQ feedback bits must be transmitted and processed within a short feedback window—on the order of few milliseconds (depends on the implementation and device though)—to meet NR’s low-latency requirements, which leaves little room for cryptographic overhead. Even appending a minimal 16–32 bit integrity field (if feasible) could significantly increase the per-packet overhead, particularly since HARQ operates in rapid, repeated cycles. Apart from significant design modifications, hardware constraints further complicate this approach, as UEs must compute or verify any authentication field (e.g., a lightweight MAC) in near real-time, risking missed timing deadlines if cryptographic operations are too slow.

From a resource standpoint, HARQ feedback typically has only a few bits for signaling ACK/NACK bursts. Extending it to include cryptographic information might crowd out existing fields or require additional sidelink symbols, cutting into spectral efficiency. In addition, because HARQ processes occur repeatedly with each transmission block, even a modest increase in per-feedback processing can accumulate, raising UE power consumption and potentially lowering throughput if the UE or network must account for these extra checks. A possible intermediate solution would be to rely on physical layer anomaly detections (e.g., verifying consistent transmission power, scrambling patterns, or channel estimates from the legitimate UE), that are valid only within a strict bound time window. 

While physical-layer validation of parameters, such as transmission power and modulation consistency, and use of directional antennas could be effective in detecting anomalous ACK/NACK signals, timing constraints for HARQ feedback are still extremely tight, often within a few of milliseconds window. While not as robust as full digital signatures (due to potential false positives/negatives), these approaches could help maintain real-time performance better than time-consuming cryptographic operations without compromising reliability in NR V2X environments as much. Nevertheless, such measures need to tested under realistic V2X scenarios to determine their robustness, and their potential advantages.

\textbf{Securing PC5 Messages.} Completely encrypting or authenticating pre-security-establishment PC5 messages can be impractical due to design changes, limited overhead budgets and the need for rapid session initiation in sidelink Mode 2. Instead, applying robust verifications—e.g., through timestamps, short sequence numbers, or stateful connection checks—can catch replayed or conflicting messages at relatively low overhead. These measures involve maintaining lightweight state on each UE (e.g., tracking recent message IDs), which adds memory and processing cost but remains significantly less demanding than full cryptographic protection.

At the same time, traffic pattern analysis (e.g., verifying that message frequencies align with known V2X protocols) imposes additional computational overhead, especially in high-density scenarios where each UE sees numerous sidelink exchanges. However, such analysis could be integrated into existing MAC or RRC procedures with minimal modifications, providing a feasible way to detect anomalies without large cryptographic fields or repeated key negotiations. In dense vehicular networks, each UE must ensure that any extra checks do not inflate connection setup times beyond acceptable bounds—particularly if the sidelink interface is used for safety-critical messages. By combining these detection methods, UEs can reduce the likelihood of MitM or malicious attachment attacks while keeping the per-message overhead relatively small.

\textbf{Preventing UE Tracking.} Frequently changing Layer-2 identifiers (e.g., via the Direct Link Identifier Update procedure) is a crucial step in thwarting adversarial tracking. However, each identifier update generates additional signaling overhead—both in the link-layer control plane (e.g., updating mapping tables) and in the application layer (if connections must be re-established). In high-traffic NR V2X environments, performing these updates too often can lead to noticeable latency spikes, as UEs must temporarily pause or reconfigure ongoing transmissions to synchronize the new identifiers among peers. Moreover, cryptographically secure randomization of each new identifier requires on-device generation of random numbers, which may be hardware-accelerated or might rely on the UE’s CPU, thus potentially affecting battery life and throughput if done at short intervals.

For groupcast or broadcast-based services, using shared group identifiers can hide individual UE identities but could reduce the precision of certain procedures (e.g., selective HARQ or targeted resource allocation). This trade-off may increase collision risk or complicates error recovery, especially as the network or autonomous Mode 2 relies on acknowledging specific UEs’ receptions. Consequently, it is currently unclear how this measure can be precisely implemented and in which use cases shared group identifiers can be used to protect the sidelink network.

Consequently, we could adopt an intermediate approach—where the UE employs moderately timed privacy timers (e.g., tens of seconds) combined with partial randomization—strikes a compromise, limiting the exposure window while keeping overhead manageable for real-time vehicular operations. As already mentioned though, the lack of specific design instructions and directives (design deficiencies) open the room for implementation flaws. The current 3GPP specifications do not establish secure generation and management procedures of such identifiers, let alone evaluating a potential trade-off between security and performance.

\textbf{General Protocol Enhancements.} Enforcing stricter security policies—such as eliminating NULL ciphers for ordinary sidelink communications and mandating authentication over the PC5 interface—can significantly reduce exploitability and is significantly less impactful on performance compared to the aforementioned measures. Nonetheless, we should keep in mind that: (1) these countermeasures do not solve the previous security flaws at the physical and MAC layers, and (2) in extremely time-sensitive V2X contexts, authentication, ciphering and integrity-protection can still cause delays in session initiation and communication and burden UEs with more frequent cryptographic operations, potentially impacting real-time performance. 

By logging security-related events (e.g., suspicious message sequences, repeated failed integrity checks), UEs can detect and respond to incidents more effectively. However, storing logs in high-throughput vehicular environments may require on-device memory, and analyzing them in real-time can consume processing cycles, implying a trade-off between thorough incident tracing and maintaining low latency. Similarly, the ability to quarantine or isolate suspicious UEs demands either network coordination or robust local procedures. Nonetheless, such methods are crucial for long-term resilience: once a malicious UE is identified, promptly notifying relevant parties (e.g., a back-end security server or the local cluster of vehicles) can avert widespread disruption. 

While these protocol enhancements may impose extra overhead and complexity, they help against the cause of sidelink vulnerabilities stemming from insufficient security defaults and permissive configuration options. Nonetheless, more investigation is needed in order to determine their practicality in real V2X scenarios.

\begin{table*}[!t]
\centering
\caption{SCI format 2-A fields for NR V2X with potential security implications. This message may be used for additional HARQ manipulation, even though HARQ relies on PSFCH for ACK/NACK signaling.}
\label{tab:sci2a-format}
\renewcommand{\arraystretch}{1.3}
\small
\begin{tabularx}{\textwidth}{l l X}
\toprule
\textbf{Parameter} & \textbf{Bit Length} & \textbf{Attacker Manipulation \& Relevance} \\
\midrule

\textbf{HARQ Process Number} &
4 bits &
Identifies the HARQ process for the current data block. Spoofing might confuse the transmitter about which HARQ process is active, potentially causing retransmission misalignment. \\

\midrule

\textbf{New Data Indicator (NDI)} &
1 bit &
Signals if the current TB is a new transmission or a retransmission. Forging this bit could mislead the receiver into treating packets incorrectly (e.g., discarding a new TB or expecting old data). However, this is \emph{not} the ACK/NACK feedback. \\

\midrule

\textbf{Redundancy Version (RV)} &
2 bits &
Specifies which redundancy version (out of 4) is used if it is a retransmission. An attacker manipulating RV could corrupt the receiver’s decoding process, although it mainly impacts HARQ efficiency rather than directly blocking resources. \\

\midrule

\textbf{Source ID} &
8 bits &
Indicates the UE sending the transport block. Spoofing could impersonate or conflate multiple sources, enabling replay or identity-based confusion. \\

\midrule

\textbf{Destination ID} &
16 bits &
Indicates the target UE/group. \\

\midrule

\textbf{HARQ Feedback Enabled/Disabled Indicator} &
1 bit &
Tells whether HARQ feedback (ACK/NACK) is expected. Spoofing “disabled” could trick the transmitter into not waiting for feedback, losing reliability. Spoofing “enabled” could cause the transmitter to expect absent feedback and force timeouts. \\

\midrule

\textbf{Cast Type Indicator} &
2 bits &
Specifies whether the sidelink transmission is unicast, groupcast, or broadcast (per Table 8.4.1.1-1 in~\cite{3gpp.38.212}). Faking cast type may lead to unexpected reception behaviors or disrupt group membership filters. \\

\midrule

\textbf{CSI Request} &
1 bit &
Requests channel state information from the receiver. An attacker toggling this bit might prompt unnecessary overhead or hamper link adaptation if the legitimate transmitter/receiver rely on accurate CSI feedback. \\

\bottomrule
\end{tabularx}
\end{table*}

%\section*{Acknowledgments}

\end{document}